\documentclass[useAMS,usenatbib]{mnras}
\usepackage{amsmath}
\usepackage{mathtools}
\usepackage{amssymb}
\usepackage{graphicx}
\usepackage{multirow}
\usepackage{color}
\usepackage{array}
\usepackage{graphicx}
\usepackage[export]{adjustbox}

\title[Parameter dependency on the reverberation models]{Parameter dependency on the public X-ray reverberation models {\sc kynxilrev} and {\sc kynrefrev}}

\author[K. Khanthasombat et al.]{K. Khanthasombat$^{1}$\thanks{E-mail: \href{mailto:M6400248@g.sut.ac.th}{M6400248@g.sut.ac.th}}, P. Chainakun$^{1,2}$\thanks{E-mail: \href{mailto:pchainakun@g.sut.ac.th}{pchainakun@g.sut.ac.th}}, A. J. Young$^3$ \\
$^1$School of Physics, Institute of Science, Suranaree University of Technology, Nakhon Ratchasima 30000, Thailand\\
$^2$Centre of Excellence in High Energy Physics and Astrophysics, Suranaree University of Technology, Nakhon Ratchasima 30000, Thailand\\
$^3$H. H. Wills Physics Laboratory, Tyndall Avenue, Bristol BS8 1TL, UK}
\date{Accepted XXX. Received YYY; in original form ZZZ}

\pubyear{2021}

\begin{document}
\label{firstpage}
\pagerange{\pageref{firstpage}--\pageref{lastpage}}
\maketitle

\begin{abstract}
We present a comparative study of the constrained parameters of active galactic nuclei (AGN) made by the public X-ray reverberation model {\sc kynxilrev} and {\sc kynrefrev} that make use of the reflection code {\sc xillver} and {\sc reflionx}, respectively. By varying the central mass ($M_{\rm BH}$), coronal height ($h$), inclination ($i$), photon index of the continuum emission ($\Gamma$) and source luminosity ($L$), the corresponding lag-frequency spectra can be produced. We select only the simulated AGN where their lag amplitude ($\tau$) and $M_{\rm BH}$ follow the known mass-scaling law. In these mock samples, we show that $\tau$ and $h$ are correlated and can possibly be used as an independent scaling law. Furthermore, $h$ (in gravitational units) is also found to be positively scaled with $M_{\rm BH}$, suggesting a more compact corona in lower-mass AGN. Both models reveal that the coronal height mostly varies between $\sim 5$--$15~r_{\rm g}$, with the average height at $\sim 10~r_{\rm g}$ and can potentially be found from low- to high-mass AGN. Nevertheless, the {\sc kynxilrev} seems to suggest a lower $M_{\rm BH}$ and $h$ than the {\sc kynrefrev}. This inconsistency is more prominent in lower-spin AGN. The significant correlation between the source height and luminosity is revealed only by {\sc kynrefrev}, suggesting the $h$--$L$ relation is probably model dependent. Our findings emphasize the differences between these reverberation models that raises the question of biases in parameter estimates and inferred correlations.

\end{abstract}


\section{Introduction}

Active galactic nuclei (AGN) are the compact, high-energy regions at the centres of galaxies that are powered by accretion of gas onto the central supermassive black hole. Matter in the surrounding areas are influenced by the strong gravitational field, hence forming an accretion disc. In the standard accretion disc scenario, the disc itself emits the optical/UV photons \citep{Shakura1973}. These photons are up-scattered by the high-energy electrons inside the corona, boosting the energy of the seed photons up to X-rays \citep{George1991, Fabian2000}.

The coronal X-rays that illuminate the disc are reprocessed via, e.g., photoionization, fluorescence, Auger process and resonant
trapping, and Compton scattering, whose resulting features are imprinted in the reflection spectra \citep{Fabian2000, Reynolds2003}. The reflection X-rays lead to a time delay, or reverberation lags, as they travel a longer distance to the observer compared to the direct continuum X-rays. The reverberation lags are then related to the distance between the corona and the accretion disc, hence can be used to probe the geometry and properties of the system \citep{Uttley2014, Cackett2021}. \cite{McHardy2007} reported the first hint of the X-ray reverberation taking place in the AGN Ark~564. The first robust detection of X-ray reverberation was in the AGN 1H~0707-495, where the high-frequency (short timescales) soft lags of $\sim 30$~s between the light curves in the reflection-dominated and continuum-dominated bands were discovered \citep{Fabian2009}. Thereafter, the reverberation lags have been discovered in several AGNs, for example, MCG-6-30-15 and Mrk 766 \citep{Emmanoulopoulos2011}, RE J1034+396 \citep{Zoghbi2011} and Ark 564 \citep{Kara2013}, increasing the amount of data so the parameter relations can be systematically analyzed.

\cite{DeMarco2013} investigated the relation between the black hole mass and the amplitude of the X-ray reverberation lags measured in the soft (Fe-L) band. They found that the lags scale with the mass. This scaling relation derived in the Fe-L band can also extrapolate to lower-mass AGNs, suggesting that the accretion process is independent of the central mass \citep{Mallick2021}. \cite{Kara2016} performed a global look on the discovered X-ray reverberating AGN and reported that the lags measured in the Fe-K band were also correlated with the mass. The lag-mass scaling relation was further confirmed by \cite{Hancock2022}, where the covering fraction that was possibly induced by the motion of non-uniform orbiting clouds was also found to be inversely correlated with the photon index and the flux of the X-ray continuum. 

Modelling of the reverberation lags has been performed independently under different geometries such as the lamp-post coronal source \citep{Cackett2014, Emmanoulopoulos2014, Chainakun2015, Epitropakis2016, Caballero2018, Ingram2019} and the extended corona environment \citep{Wilkins2016, Chainakun2017, Chainakun2019, Hancock2023,Lucchini2023}. Meanwhile, the public models for the X-ray reverberation lags, {\sc kynxilrev} and  {\sc kynrefrev}\footnote{\url{https://projects.asu.cas.cz/stronggravity/kynreverb}} models, were developed \citep{Caballero2018} based on the {\sc kyn} model\footnote{\url{https://projects.asu.cas.cz/stronggravity/kyn}} previously developed by \cite{Dovciak2004a} and \cite{Dovciak2004b}. The {\sc kynxilrev} and {\sc kynrefrev} models compute the reverberation time lags under the lamp-post assumption by using the ionized X-ray reflection from the {\sc xillver} \citep{Garcia2010, Garcia2013,Garcia2014} and {\sc reflionx} \citep{Ross1999,Ross2005} table models, respectively. Thus far, {\sc xillver} and {\sc reflionx} have been the most widely-used X-ray reflection models \citep{Bambi2021}. 

While both reflection models are generally comparable, the {\sc xillver} model uses more updated atomic composition data of the accretion disc in calculating the photon-disc interaction and corresponding X-ray reflection. The main differences between the spectra from both models are likely in the soft ($\lesssim 1$~keV) band, especially when the photon index of the X-ray continuum is low \citep{Garcia2013}. Instead of focusing on the time-average spectrum, the aim of this work is to investigate the differences between the {\sc kynxilrev} and {\sc kynrefrev} models in explaining the X-ray timing data. The mock AGN samples are produced following the current trend of the lag-mass scaling relation as reported by \cite{DeMarco2013}. The results are compared between those simulated from the {\sc kynxilrev} and {\sc kynrefrev} models. We test whether the results from both models are consistent in determining the key parameters of the system such as the coronal height and luminosity. 

More specific details of the {\sc kynxilrev} and {\sc kynrefrev} models as well as the parameters investigated here are described in Section 2. Section 3 presents how the lag-frequency spectra are generated and further selected by matching the observational trend of the lag-mass scaling relation. In Section 4, we investigate the obtained parameters and existing correlations that can be derived when using {\sc kynxilrev} and {\sc kynrefrev} models, and highlight the differences between the results from both models. The discussion and conclusion towards the model dependence is given in Section 5.

\section{X-ray reverberation models}

{\sc kynxilrev} and {\sc kynrefrev} is the X-ray reverberation coding model related to the {\sc kyn} package \citep{Dovciak2004a, Dovciak2004b, Caballero2018} that can calculate the frequency-dependent time lags by using the energy-integrated X-ray spectrum from the {\sc xillver} \citep{Garcia2010,Garcia2013} and {\sc reflionx} (\cite{Ross1999, Ross2005}) models. We use \texttt{reflionx.mod} as the {\sc reflionx} table model, and specifically employ \texttt{xillverD-5.fits} for the {\sc xillver} reflection model. 

%, where the X-ray reprocessing on the accretion disc is calculated following the photoionisation routines from {\sc xstar} code \citep{Kallman2001}. 

The incident angle of the X-rays irradiating the disc has a significant impact on the rest-frame reflection spectrum. While {\sc reflionx} assumes an isotropic point source situated above a semi-infinite slab of cold gas, {\sc xillver} assumes that the coronal X-rays illuminate the disc with an incident angle of 45 degrees (see, e.g., \citealt{Dauser2013} and discussion in \citealt{Bambi2021}). Furthermore, most of the solar abundances of the elements (e.g., C, O, Ne, Mg and Fe) used in {\sc xillver} are lower than those in {\sc reflionx} \citep{Garcia2013}. This is because {\sc reflionx} adopted the elemental solar abundances from \cite{Morrison1983}, while the {\sc xillver} model calculated the X-ray reprocessing on the accretion disc following the photoionisation routines from {\sc xstar} code \citep{Kallman2001}, which adopted the abundances of the elements from \cite{Grevesse1998}. Among the widely-utilized reflection models, {\sc xillver} is probably the most sophisticated one. The {\sc xstar} code incorporated by {\sc xillver} contains the most comprehensive atomic database for the detailed treatment of the radiative transfer to model photoionized X-ray reflection spectra. {\sc xillver} also enhances, in particular, the thorough calculation of the K-shell photoabsorption of prominent ions such as N and O \citep{Kallman2004, Garcia2005, Garcia2009, Garcia2013, Bambi2021}.

The {\sc kynxilrev} and {\sc kynrefrev} models are based on the assumption of the lamp-post coronal geometry (i.e. the corona is an isotropic point source located on the rotational axis of the black hole and produces a flash power-law radiation). The accretion disc is given to be optically thick and geometrically thin that obeys the standard model of \cite{Novikov1973}. The inner edge of the accretion disc is set to be equivalent to the innermost stable circular orbit (ISCO) while the outer edge is fixed at $1000~r_{\rm g}$. Note that throughout this work, the geometrical units of distance and time of $r_{\rm g} = GM_{\rm BH}/{c^2}$ and $t_{\rm g} = GM_{\rm BH}/{c^3}$ are used, where $G$ is the gravitational constant, $M_{\rm BH}$ is the black hole mass and $c$ is the speed of light. The disc has a constant density profile, while the ionization state of the disc is allowed to vary with the incident flux. The X-ray continuum is given in terms of a cut-off power law with the photon index $\Gamma$. The high-energy cut-off is fixed at 300 keV. 

The frequency-dependent Fe-L lags are calculated using the energy bands of 0.3--0.8 and 1--4~keV. In the standard Fourier technique \citep{Nowak1999}, the cross-spectrum $C(f)=S^{*}(f)H(f)$ is firstly computed, where $S(f)$ and $H(f)$ are the Fourier forms of the soft and hard band light curves. The symbol $^{*}$ represents the complex conjugate of the soft component. Next the lags are computed via the argument of the cross-spectrum (i.e. the phase difference) in the Fourier space: $\tau = {\rm arg}\;C(f)/(2 \pi f)$. Note that the phase difference is defined over $-\pi$ to $\pi$, so the phase wrapping occurs and the lag profile begins to fluctuate around 0 at the frequencies of $f \geq 1/2\tau$. Furthermore, we set the parameter $xsw=16$ (default value) in both {\sc kynxilrev} and {\sc kynrefrev} models, meaning that the rebinning is done in real and imaginary parts before the lags are computed and further diluted by the primary spectra with respect to the given (observed) energy band. 

The spin parameter of the black hole, if not stated, is fixed at $a=1$. The free parameters investigated here are the black hole mass ($M_{\rm BH}$), the coronal height ($h$), the inclination ($i$), the photon index of the primary continuum ($\Gamma$) and the observed 2--10~keV luminosity in the units of the Eddington luminosity ($L/L_{\rm Edd}$). We vary $M_{\rm BH}$ in the range of ${10^5}$--${10^9 M_\odot}$, which also covers the low mass AGN as investigated in \cite{Mallick2021}. The coronal height $h$ is varied in the range of 2--30~$r_{\rm g}$, while the inclination $i$ is between 15$^{\circ}$--75$^{\circ}$. The photon index $\Gamma$ and the luminosity $L$ are varied between 1.8--3.0 and 0.001--0.5 $L_{\rm Edd}$, respectively. These parameter ranges are presented in Table~\ref{tab:parameter}. Note that we focus on the case when the inner disc is fixed at ISCO, but we also try to vary the inner radius (the result is presented in Appendix~\ref{sec:a1}). The \texttt{rand()} is used to generate the random integer with the seed generated by \texttt{srand(time(0))} so that it returns a different sequence of random numbers each time it is executed.

%The combination of \texttt{rand()} and \texttt{srand(time(0))} functions are used in order to uniformly generate the random values within the range of each parameter. 

\begin{table}
 \caption{List of parameters and the random distribution used in the time lag simulation process. Note that the luminosity is measured in the 2--10~keV energy band. Uniform-lin and uniform-log mean that the parameter values are randomly selected from a uniform probability density function in the linear and logarithmic space, respectively.}
 \label{tab:parameter}
 \begin{tabular}{lcc}
  \hline Parameter & Distribution, Range & Unit  \\ [5pt]
  \hline
  \\ [\dimexpr-\normalbaselineskip+2pt]
  Black hole mass (${M_{\rm BH}}$) & Uniform-log, $10^5$--$10^9$ & $M_\odot$ \\ [2pt]
  Corona height ($h$)              & Uniform-lin, 2--30         & ${r_{\rm g}}$ \\ [2pt]
  Inclination ($i$)                & Uniform-lin, 15--75        & degrees \\ [2pt]
  Photon index ($\Gamma$)          & Uniform-lin, 1.8--3.0 \\ [2pt]
  Luminosity ($L$)                 & Uniform-log, 0.001--0.5    & ${L_{\rm Edd}}$\\ [2pt]
  \hline
 \end{tabular}
\end{table}

\section{Lag-mass scaling relation and data simulations}

\begin{figure}
 \includegraphics[width=\columnwidth]{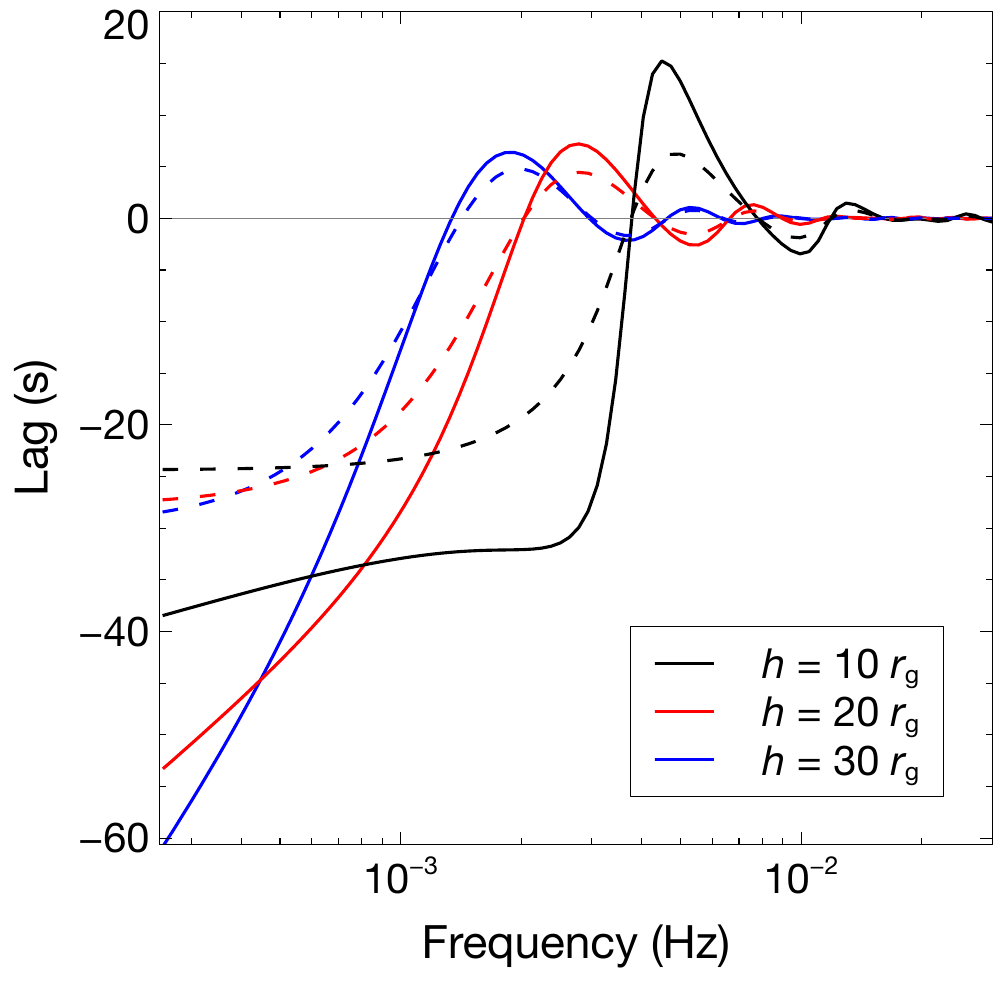}
 %\put(-70,100){$M_{\rm BH} = 10^{6} M_\odot$}
 \vspace{-0.5cm}
 \caption{Examples of the lag-frequency spectra from the {\sc kynxilrev} (dashed lines) and {\sc kynrefrev} (solid lines) models when the coronal height is varied to be $h=10$, 20 and 30~$r_{\rm g}$. The lags are calculated using the 0.3--0.8 (soft) and 1--4 keV (hard) energy bands. Negative lags indicate the soft band is lagging the hard band. The black hole mass is fixed at $M_{\rm BH} = 10^{6} M_\odot$. The other AGN parameters are $a =1$, $i =30^{\circ}$, $\Gamma =2$ and $L = 0.001 L_{\rm Edd}$}
 \label{fig:lag-freq_spectrum}
\end{figure}

The studies on the observational surveys of the reverberation lags revealed that the AGN with a higher mass seems to exhibit larger reverberation lags while the observed frequencies of the lags are lower \citep[e.g.][]{DeMarco2013, Kara2016, Hancock2022}. Here, the lag-frequency spectra in the Fe-L band of the mock AGN samples are generated using the {\sc kynxilrev} and {\sc kynrefrev} models, by varying the parameters shown in Table~\ref{tab:parameter}. The examples of the lag-frequency spectra simulated from both models are presented in Fig.~\ref{fig:lag-freq_spectrum}. The negative lags are defined as the soft band lagging behind the hard band. As expected, the maximum amplitude of the lag increases with the coronal height. The phase wrapping also occurs at lower frequencies for higher source height. Furthermore, it can be seen that the phase-wrapping frequencies from both models are quite consistent, but, for each source height, the lag amplitude of the {\sc kynxilrev} model is smaller than that from the {\sc kynrefrev}. This is expected since the reflection spectrum from the {\sc xillver} model is more absorbed at low energies \citep{Garcia2013}, so the soft lags from the {\sc kynxilrev} is more diluted than the {\sc kynrefrev} model. The dilution effects reduce the amplitude of the lags without affecting the phase wrapping \citep[see][for discussion on intrinsic lags and dilution]{Wilkins2013, Kara2014, Chainakun2023}. 

From the mass-scaling law, we know the lag amplitude ($\tau_{\rm obs}$) and the particular frequency ($\nu_{\rm obs}$) at which we expect to see the lags for a given black hole mass. Therefore, the simulated lag spectra are screened in order to check if they are consistent with the mass-scaling law suggested by \cite{DeMarco2013}:
\begin{equation}
\log \tau_{\rm obs} = 1.98(\pm0.08)+0.59(\pm0.11) \: {\rm log}(M_7) \: ,
    \label{eq:lag-mass}
\end{equation}
\begin{equation}
{\rm log} \ \nu_{\rm obs} = -3.50(\pm0.07)-0.47(\pm0.09) \: {\rm log}(M_7) \: ,
   \label{eq:freq-mass}
\end{equation}
where $\tau_{\rm obs}$ is the lag amplitude, $\nu_{\rm obs}$ is the frequency where the lags are seen and $M_7$ is $M_{\rm BH}/10^7 M_\odot$.

%However, the fluctuations along the accretion disc that propagate inwards on longer timescales than those from the inner disc reflection can produce the positive hard lags at relatively low frequencies \citep{Kotov2001,Arevalo2006}. 

To sort the simulated lag amplitude ($\tau_{\rm sim}$) and frequency ($\nu_{\rm sim}$) that match $\tau_{\rm obs}$ and $\nu_{\rm obs}$, the lag spectra are binned using the bin size comparable to those typically used in the timing analysis of the AGN such as 1H0707--495 and IRAS~13224--3809 \citep{Caballero2018, Caballero2020}. Then, $\tau_{\rm sim}$ is determined from the frequency bin that is before the phase wrapping occurs. The central frequency of that bin is used as a proxy for $\nu_{\rm sim}$. Then, for a given $M_{\rm BH}$, we check for the consistency between the observed and the simulated lag values, by allowing $\sim 20$ and 2 per cent of the deviations between $\tau_{\rm sim}$ and $\tau_{\rm obs}$, and between $\nu_{\rm sim}$ and $\nu_{\rm obs}$, respectively. The effects of this allowed uncertainty are also investigated. All simulated spectra whereby phase wrapping occurs at frequencies below $\nu_{\rm obs}$ are excluded. Only the samples that are consistent with the observed mass-scaling relation are included as the mock AGN samples for further analysis.

The fluctuations along the accretion disc that propagate inwards on longer timescales than those from the inner disc reflection can produce the positive hard lags at relatively low frequencies \citep{Kotov2001,Arevalo2006}. We note that \cite{DeMarco2013} identified the value of the soft lag and the associating frequency from the amplitude and the frequency of the minimum point in the negative-lag profile, which is possible since the positive and negative lags can be clearly distinguished from the observed data. Here, we model only the negative reverberation lags so the minimum point as in \cite{DeMarco2013} cannot be easily identified (i.e. we do not know the amount of the competing positive lags and the highest frequencies where they are possibly dominated). Instead, we choose to determine the value of the soft lag from the frequency bin just before the phase wrapping to ensure that the clean reverberation signature is probed. Nevertheless, for the accepted AGN data, the frequency bin where the lag is seen must also match the frequency bin reported by \cite{DeMarco2013}. This further ensures that our mocked data have both lag amplitude and the corresponding frequency following the mass-scaling law. An uncertainty due to the effect of the positive hard lags is also discussed in Appendix~\ref{sec:a2}.

\begin{figure}
 \includegraphics[width=\columnwidth]{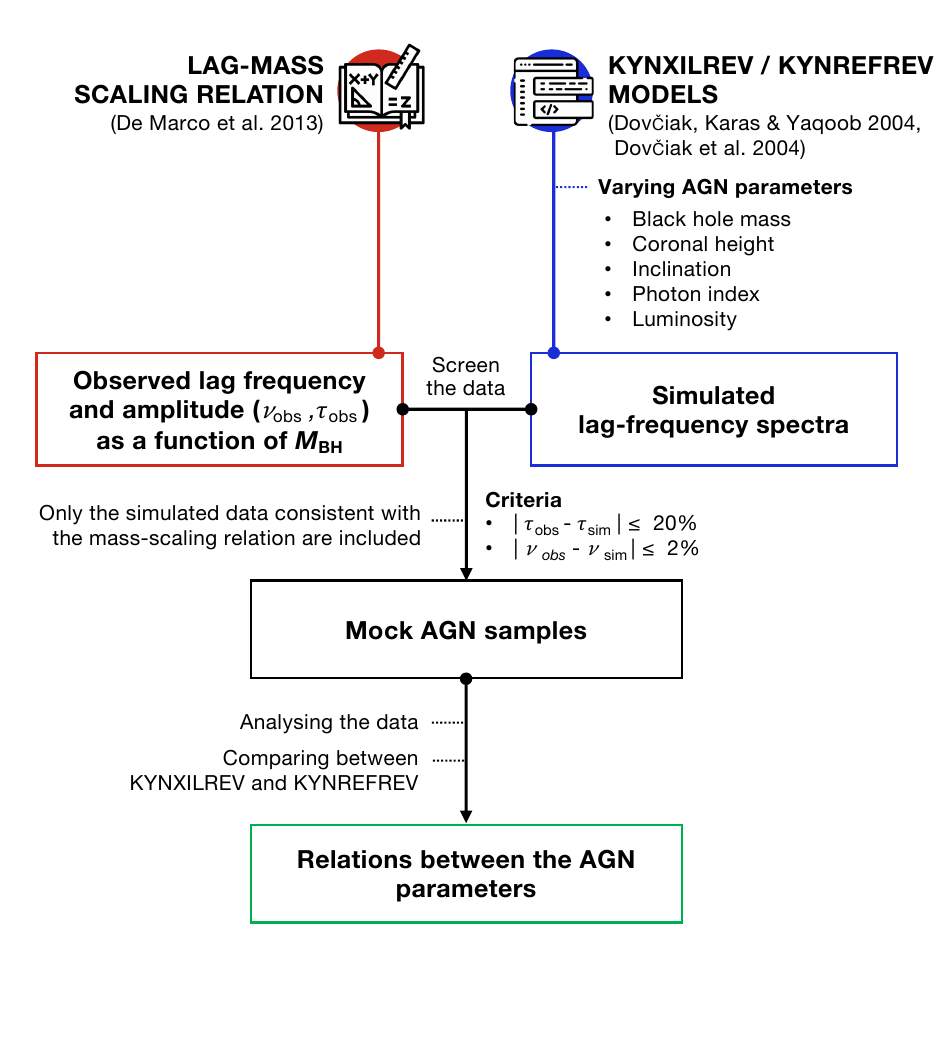}
 \vspace{-1.5cm}
 \caption{The flowchart representing the investigation process for this work.} \label{fig:workflow}
\end{figure}

The flowchart illustrating the overall workflow of this work is presented in Fig.~\ref{fig:workflow}. After the mock AGN samples that obey the mass-scaling law are obtained, the parameters derived from the {\sc kynxilrev} and {\sc kynrefrev} models are compared. While we perform 25,000 simulations of the lag frequency spectra, there are $\sim 117$ (115) of them produced from {\sc kynxilrev} ({\sc kynrefrev}) that fit in the lag-mass scaling relation and are accepted under our criteria. The correlations between the parameters implied from both models are then investigated. Finally, we allow the lag-mass scaling relation to deviate from the current trend and see its global effect on the parameters of the system.

\section{Results}

\begin{figure*}
    \centerline{
        \includegraphics[width=0.35\textwidth]{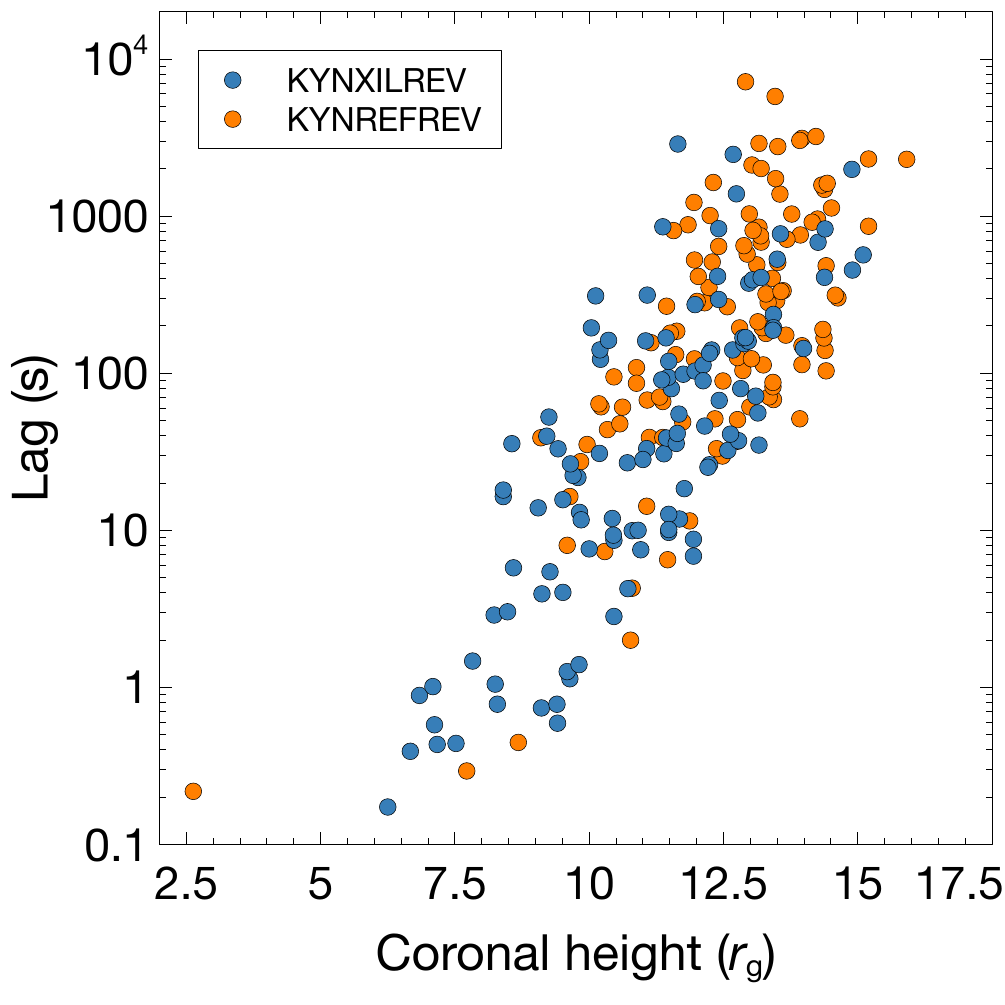}
        \includegraphics[width=0.35\linewidth]{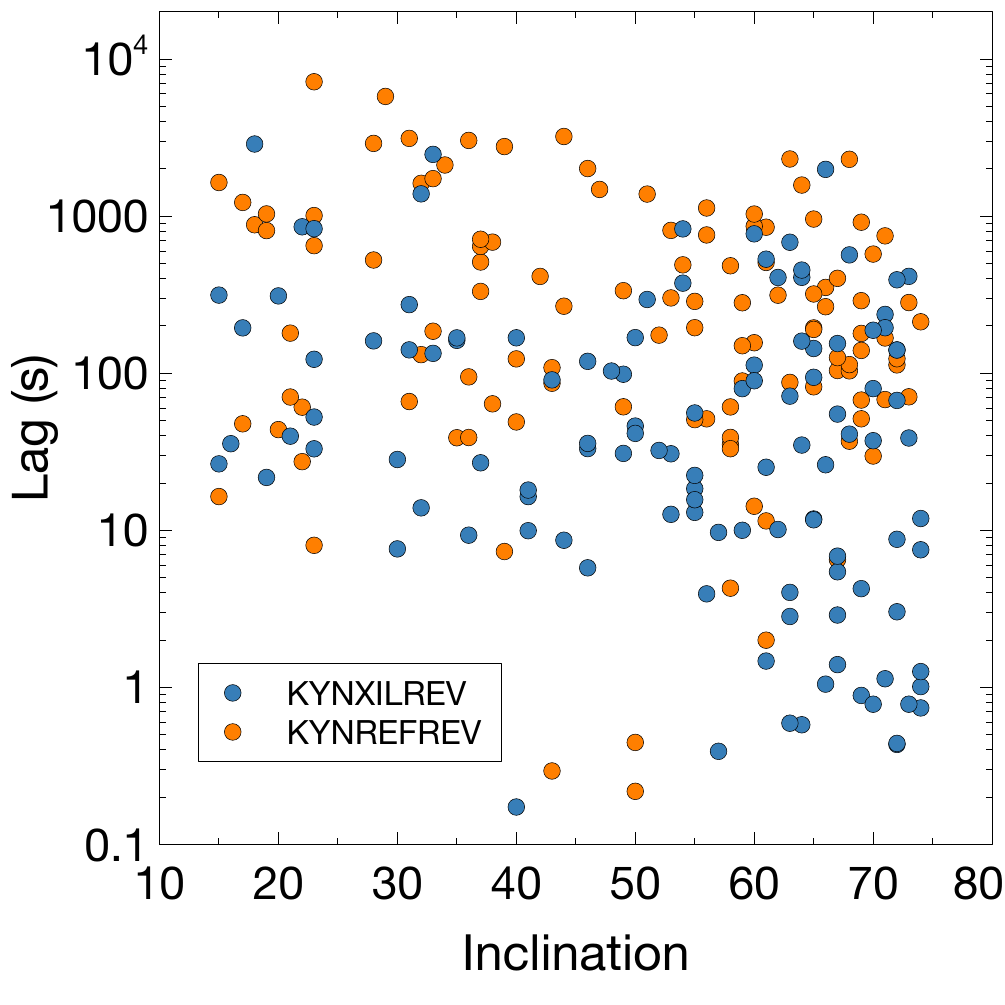}
    }   
    % \vspace{-0.05cm}
    \centerline{
        \includegraphics[width=0.35\textwidth]{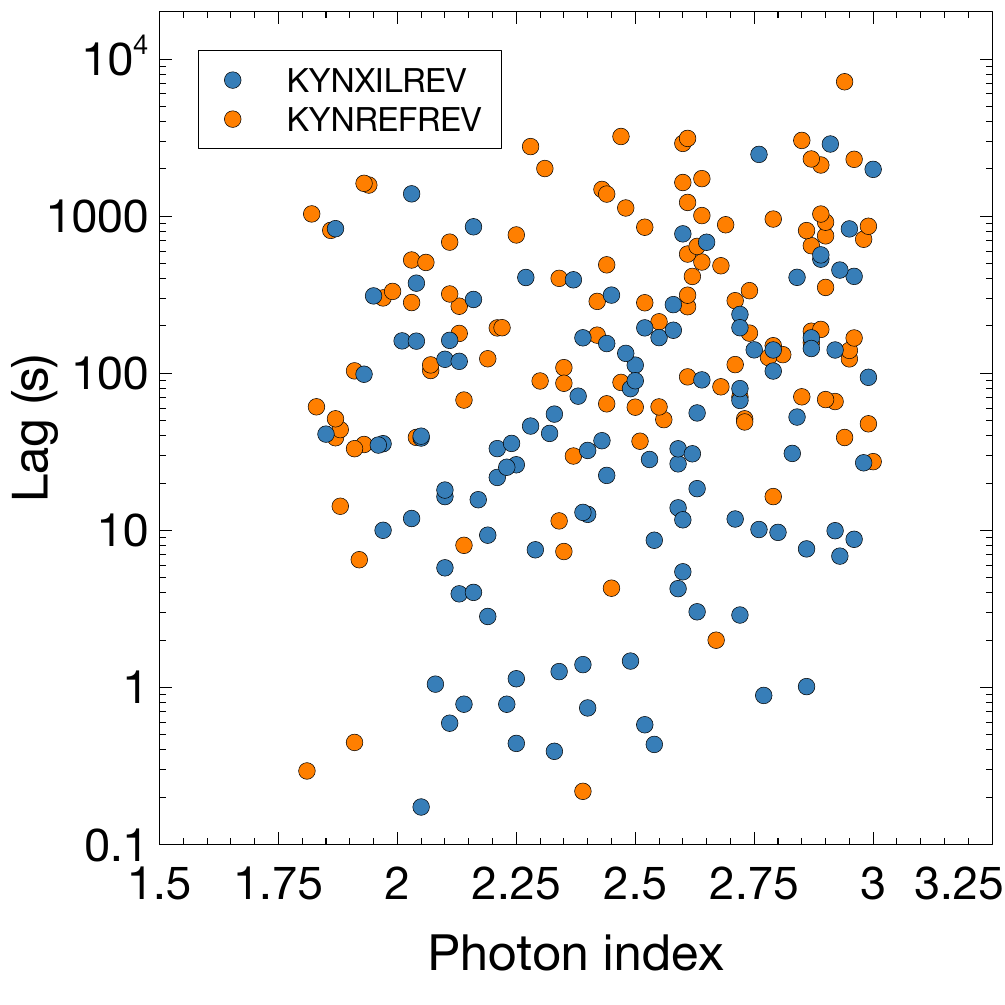} 
        \includegraphics[width=0.35\linewidth]{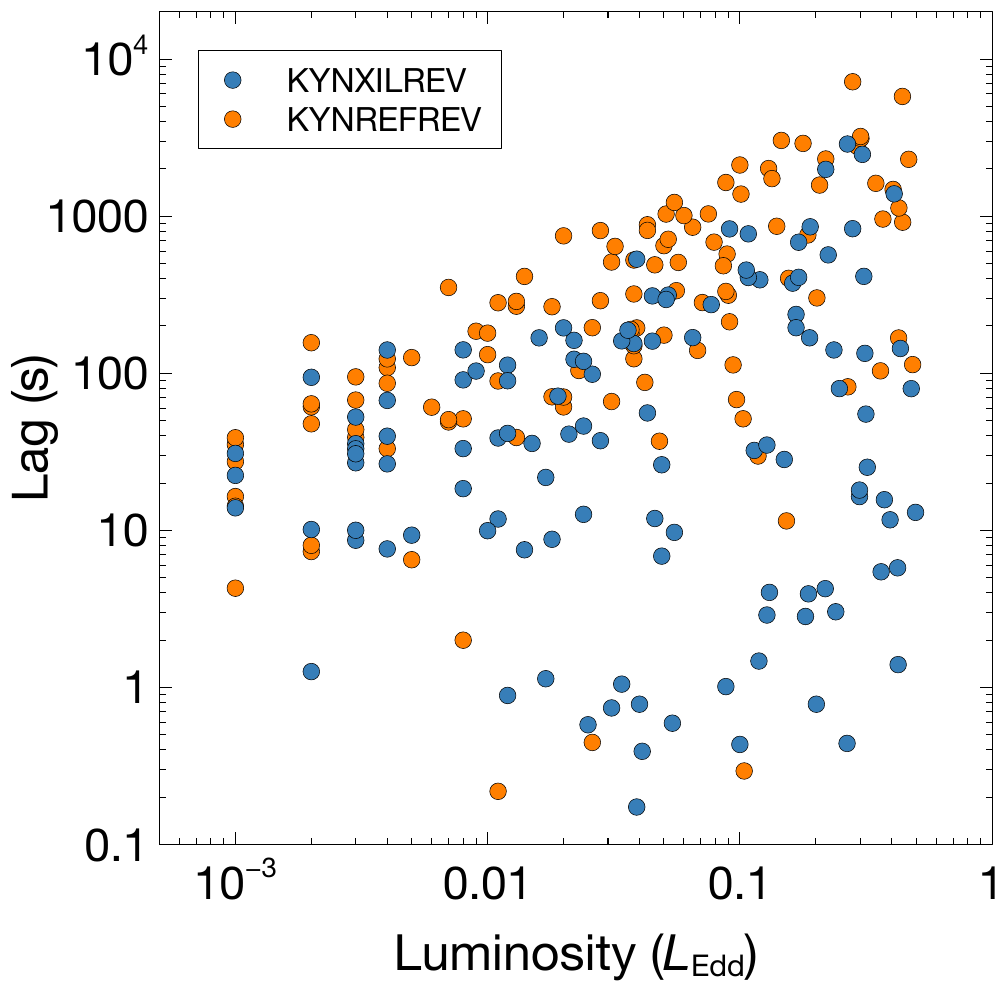}
    }   
    \vspace{-0.2cm}
    \caption{The relations between time lags and other parameters of the mock AGN samples that follow the lag-mass scaling law as obtained by {\sc kynxilrev} (blue) and {\sc kynrefrev} (orange). See text for more details.}
    \label{fig:lag_corelation}
\end{figure*}

Once the mock AGN samples that match the mass-scaling law are gathered, the relations of their obtained parameters including $\tau$, $M_{\rm BH}$, $h$, $i$, $\Gamma$ and $L$ are analyzed. Fig.~\ref{fig:lag_corelation} illustrates the relations between the observed lags and other parameters, comparing between what was derived by {\sc kynxilrev} and {\sc kynrefrev} models. The moderate monotonic correlation between $\tau$ and $h$ is significant ($p < 0.05$), with the Spearman's rank correlation coefficient of $r_{\rm s}$ = 0.39 and 0.33 for {\sc kynxilrev} and {\sc kynrefrev}, respectively. However, the {\sc kynxilrev} model suggests more solutions towards lower lags and lower source height that can still follow the mass-scaling law. The lags are found to correlate with the luminosity only when using the {\sc kynrefrev} model, where $r_{\rm s} = 0.55$. There is no correlation between the lags and inclination, and between the lags and photon index of the X-ray continuum from either {\sc kynxilrev} or {\sc kynrefrev} model.

\begin{figure*}
    \centerline{
        \includegraphics[width=0.35\textwidth]{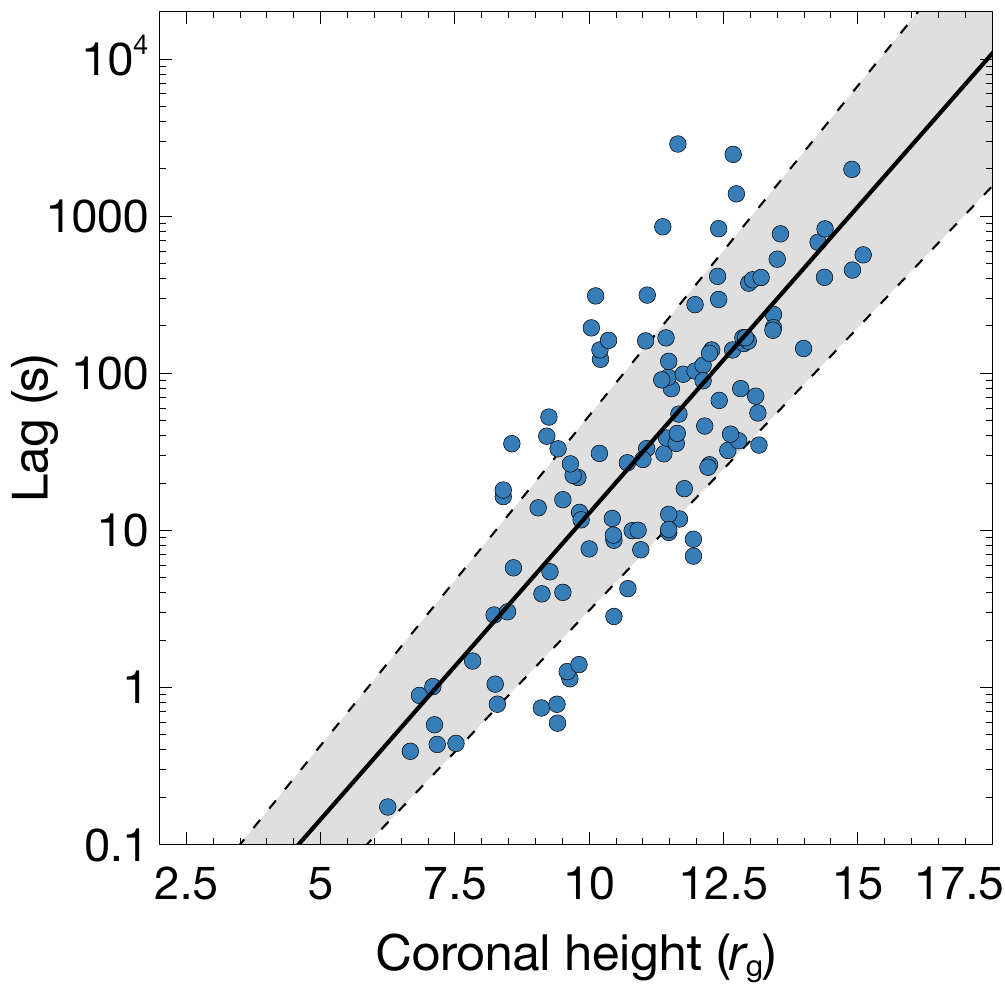}
        \put(-65,45){{\sc KYNXILREV}}
        \includegraphics[width=0.35\linewidth]{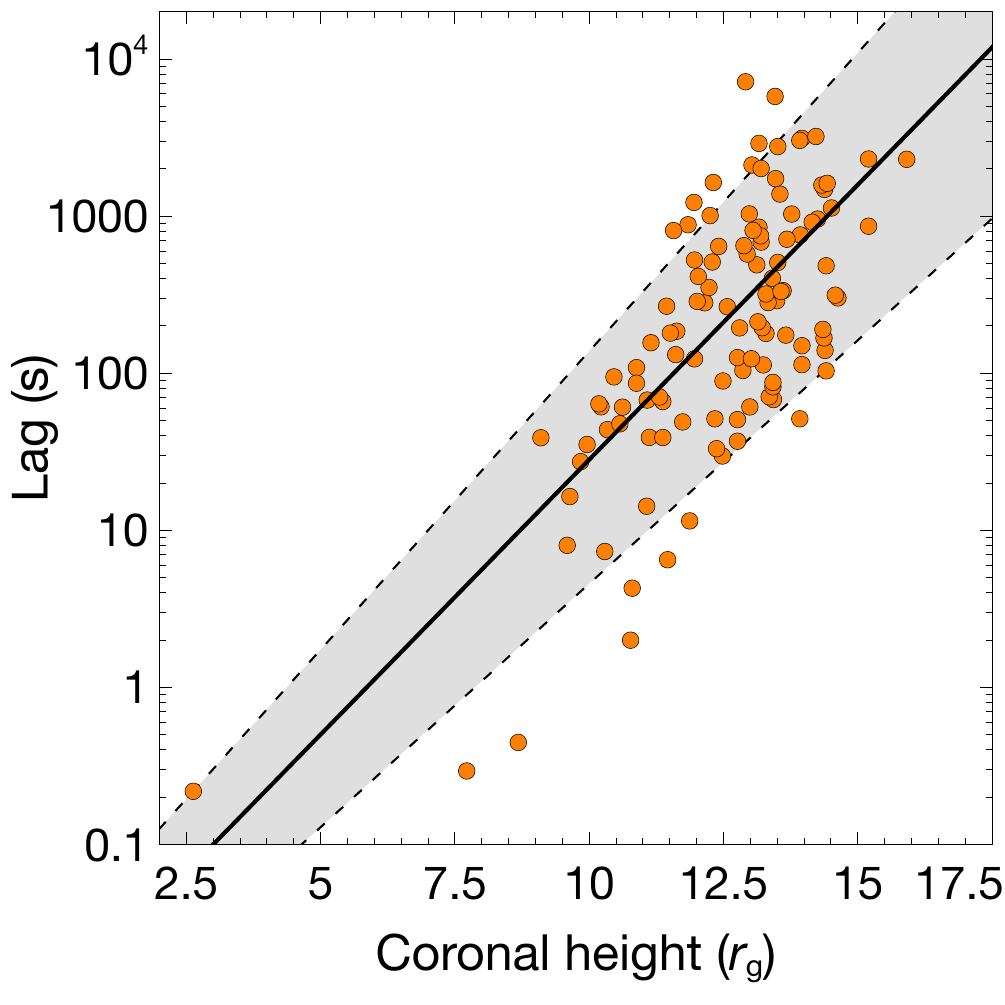}
        \put(-65,45){{\sc KYNREFREV}}
    }  
    \vspace{-0.2cm}
    \caption{The best fit to the $\tau$--$h$ relation implied from {\sc kynxilrev} (left panel) and {\sc kynrefrev} (right panel). The 1--$\sigma$ standard errors associated with the respective regression coefficients} are overplotted on the data as the grey-shaded region between dashed lines.
    \label{fig:lag-h_correlation}
\end{figure*}

The $\tau$--$h$ relation shown in Fig.~\ref{fig:lag_corelation} suggests that, in addition to the black hole mass, the observed lags also depend on the height of the corona which varies in all AGN. We then derive the $\tau$--$h$ relation by performing linear regression fitting to the data obtained from {\sc kynxilrev} and {\sc kynrefrev} models. The fitting results are presented in Fig.~\ref{fig:lag-h_correlation}. We find that the $\tau$--$h$ relation can be written in the form of
\begin{equation}
\log \tau = -2.79(\pm 0.32)+ 0.39 (\pm 0.03) \: h \: ,
    \label{eq:lag-h-xil}
\end{equation}
\begin{equation}
\log \tau = -2.05(\pm 0.39)+ 0.35 (\pm 0.03) \: h \: ,
    \label{eq:lag-h-ref}
\end{equation}
for {\sc kynxilrev} and {\sc kynrefrev} models with corresponding $R^2 =$ 0.63 and 0.53, respectively. Note that $R^2$ is the coefficient of determination that represents the variation from the true values of a dependent variable as predicted by the regression model ($R^2 = 1$ is a perfect fit). The uncertainty of each model coefficient is estimated from the 1--$\sigma$ standard error calculated by taking the square root of the diagonal elements of the obtained covariance matrix.

Then, we investigate whether the lag-mass scaling relation prefers any specific value of the black hole spin. The spin parameters are varied to be $a=0$, 0.5 and 1 for both models. Note that other parameters are uniformly varied within the range specified in Table~\ref{tab:parameter}, and only the samples that are consistent with the mass-scaling law are included in the analysis.

\begin{figure}
    \centerline{
 \includegraphics[width=0.4\textwidth]{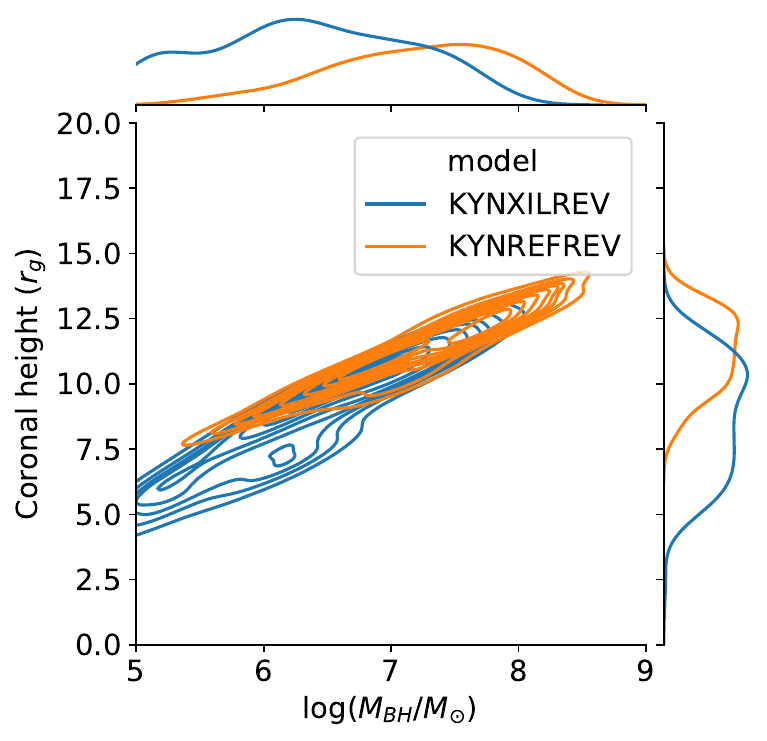}
        \put(-152,150){$a = 0$}
%\put(-131,134){$t_{\rm w}=200$~s}
    }
    % \vspace{-0.15cm}
    \centerline{
        \includegraphics[width=0.4\textwidth]{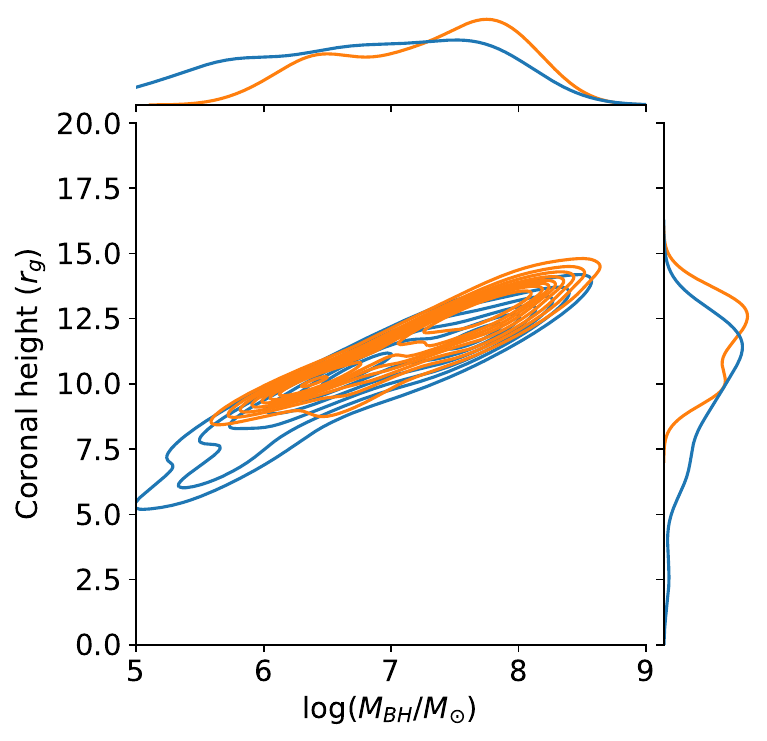}
        \put(-152,150){$a = 0.5$}
    }
        \centerline{
        \includegraphics[width=0.4\textwidth]{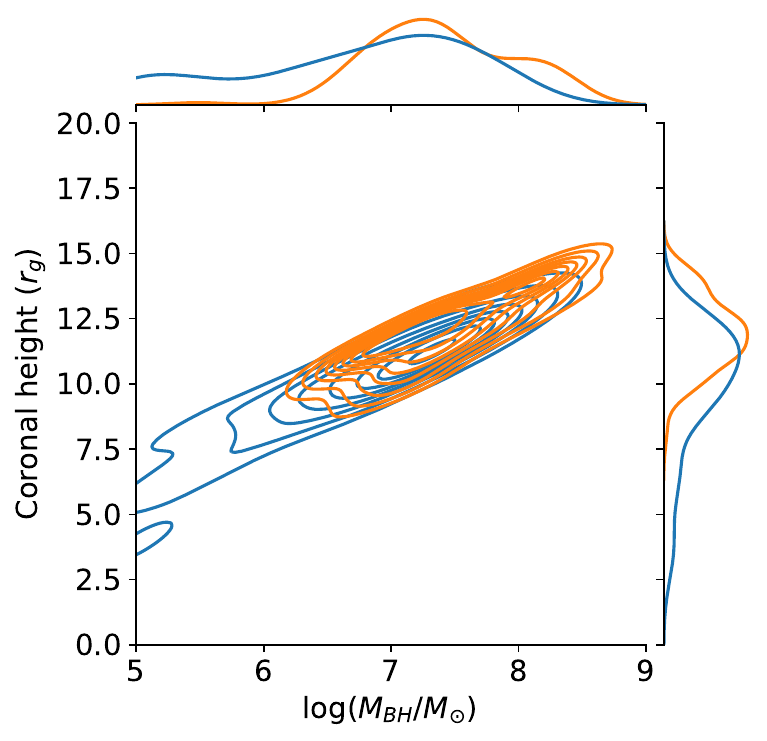}
        \put(-152,150){$a = 1$}
    }
    %\vspace{-0.2cm}
    \caption{The relationship between the coronal height and the black hole mass of the mock AGN samples for different black hole spin of $a=0$ (top panel), 0.5 (middle panel) and 1 (bottom panel). The results are generated and compared between the {\sc kynxilrev} (blue) and {\sc kynrefrev} (orange) models using the KDE. The curves outside the panels show the marginal distribution of the data for each parameter and model.
    }
     \label{fig:spin_test}
\end{figure}

Fig.~\ref{fig:spin_test} represents the $h$--$M_{\rm BH}$ distribution in the cases of low, medium, and high spin, derived from the layered kernel density estimate (KDE) in {\tt seaborn} \citep{Waskom2021}. The KDE plot illustrates the smoothed-density distribution of the data using the Gaussian kernel with the contour lines to reveal the cluster and trend of the scattered data points. Interestingly, the solutions of low, medium and high spin are all possible. Nevertheless, when $a=0$ (upper panel), the {\sc kynxilrev} provides more samples at lower masses of $M_{\rm BH} \lesssim 6.5M_{\odot}$, while the {\sc kynrefrev} model provides more samples at higher masses of $M_{\rm BH} \gtrsim 7.5M_{\odot}$. This can lead to a bias due to the choice of the model used in determining the black hole mass and the coronal height in AGN that host a low-spin black hole. Even though the preferred solutions of $h$ and $M_{\rm BH}$ for both models become more consistent in the cases of $a=0.5$ and 1, the tendency that the {\sc kynxilrev} suggests the lower mass and lower source height is still noticeable. Despite this, both models strongly suggest the correlation between $h$ and $M_{\rm BH}$, regardless of the black hole spin.

\begin{figure*}
    \centerline{
        \includegraphics[width=0.35\textwidth]{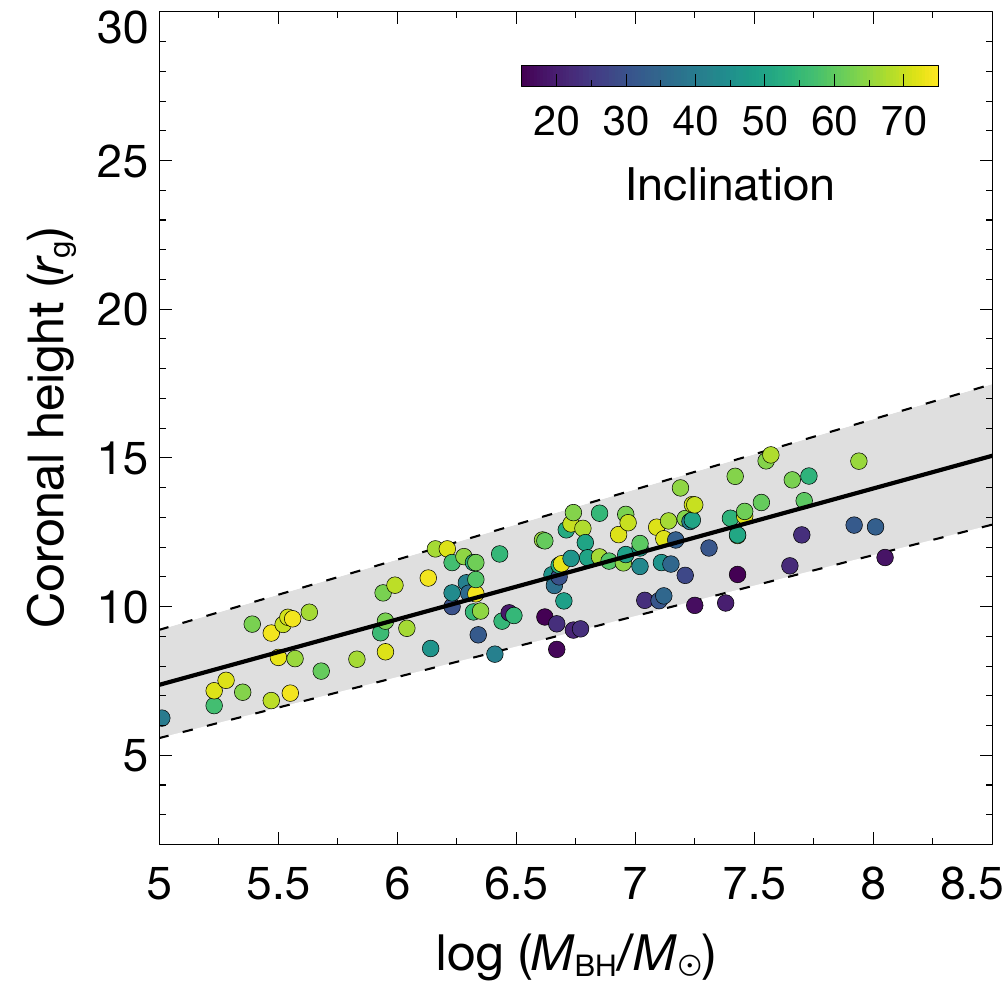}
        \put(-65,45){{\sc KYNXILREV}}
        \put(-145,160){$\Delta \tau_{\rm err} \lesssim 20\%$}
        \includegraphics[width=0.35\linewidth]{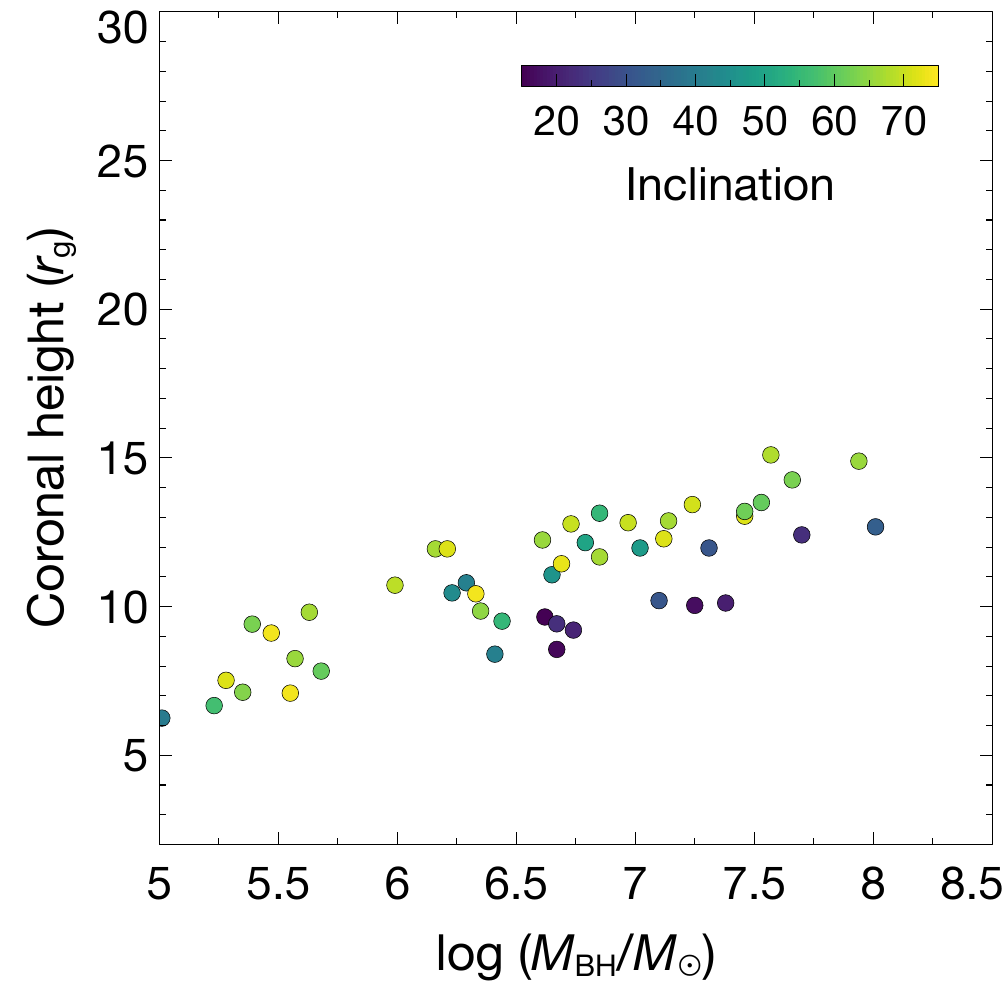}
        \put(-65,45){{\sc KYNXILREV}}
\put(-145,160){$\Delta \tau_{\rm err} \lesssim 10\%$}
    }   
    % \vspace{-0.15cm}
    \centerline{
        \includegraphics[width=0.35\textwidth]{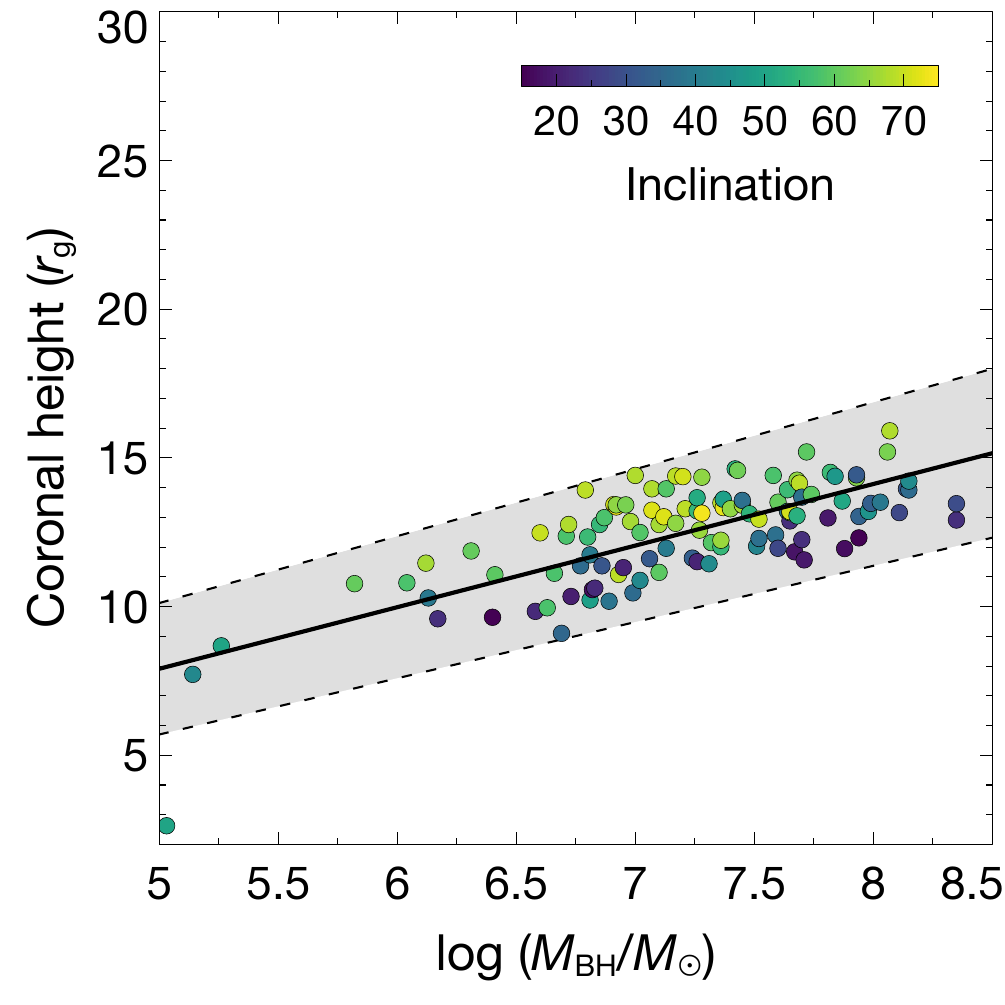} 
        \put(-65,45){{\sc KYNREFREV}}
        \put(-145,160){$\Delta \tau_{\rm err} \lesssim 20\%$}
        \includegraphics[width=0.35\linewidth]{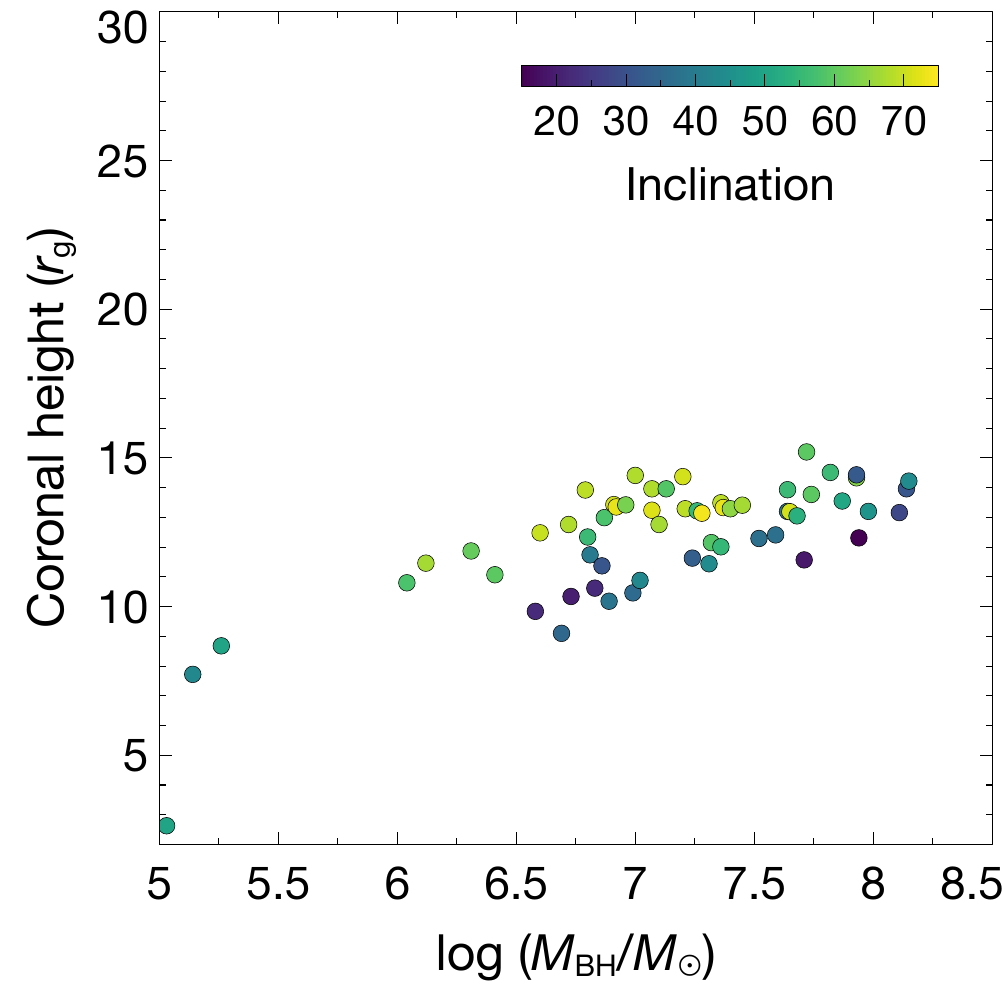}
        \put(-65,45){{\sc KYNREFREV}}
        \put(-145,160){$\Delta \tau_{\rm err} \lesssim 10\%$}
    }   
    %\vspace{-0.2cm}
    \caption{Obtained $h$--$M_{\rm BH}$ solutions from {\sc kynxilrev} (top panels) and {\sc kynrefrev} (bottom panels) models when the acceptable lag errors are limited to be within 20 (left panels) and 10 (right panels) per cent. The color indicates the value of the inclination angle (navy to yellow, $i$ = 15$^{\circ}$--75$^{\circ}$). The solid line represents the best-fit linear model. The grey-shaded region between dashed lines show the uncertainty estimated from the 1--$\sigma$ standard errors of the respective model coefficients.}
    \label{fig:height-mass}
\end{figure*}

The scatter plots for the simulated data showing the relation between $h$, $M_{\rm BH}$ and $i$ when $a=1$ are presented in Fig.~\ref{fig:height-mass}. Here, we also illustrate how the choice of the acceptable errors, $\Delta \tau_{\rm err}$, affects the results. When $\Delta \tau_{\rm err}$ decreases, the number of the solutions decreases which is expected since in doing this we accept only the mock samples that show more alignment with the observed scaling relation. By changing the threshold to accept the amplitude of the lag from 20 to 10 per cent, the number of sets of parameters produced from kynxilrev (kynrefrev) that are accepted decreases from $\sim 117$ (115) to 47 (59). In any case, both {\sc kynxilrev} and {\sc kynrefrev} show a similar trend where the corona tends to be located at a higher gravitational height for a larger $M_{\rm BH}$. Interestingly, this suggests that we can perhaps establish the $h$--$M_{\rm BH}$ scaling relation as what is done for the lag--$M_{\rm BH}$ relation. We then fit the $h$--$M_{\rm BH}$ data with a linear model. For {\sc kynxilrev} and {\sc kynrefrev} models, the best-fit $h$--$M_{\rm BH}$ relations are 
\begin{equation}
h = -3.63(\pm 1.04)+ 2.20 (\pm 0.16) \: \log (M_{\rm BH}/M_{\odot}) \: ,
    \label{eq:h-mass-xil}
\end{equation}
\begin{equation}
h = -2.44(\pm 1.31)+ 2.07 (\pm 0.18) \: \log (M_{\rm BH}/M_{\odot}) \: ,
    \label{eq:h-mass-ref}
\end{equation}
where the obtained $R^2$ is 0.64 and 0.54, respectively. However, the coronal height and the black hole mass are limited at high ends to be $h \lesssim 15~r_{\rm g}$ and $M_{\rm BH} \lesssim 10^{8.5}~M_{\odot}$. The models with higher values of the coronal height than $\sim 15~r_{\rm g}$ produce the lags which are too long and the phase-wrapping also occurs at too low frequencies that do not fit in the mass scaling relation. Note that we also investigate the case when the disc is truncated before the ISCO (see Appendix~\ref{sec:a1}).

According to Fig.~\ref{fig:height-mass}, the inclination can be expected in the complete given range of 15--75$^{\circ}$, with small inclinations (20--30$^{\circ}$) commonly found in $M_{\rm BH} \gtrsim 10^{6.5} M_{\odot}$. Although the higher inclinations ($\geq 70^{\circ}$) appear throughout the entire range of the mass for the {\sc kynxilrev}, they seem to be rarely found in {\sc kynrefrev} and only appear when $M_{\rm BH} \sim 10^{7} M_{\odot}$. The trend of the mock samples remain almost the same even when the acceptable uncertainty, e.g. $\Delta \tau_{\rm err}$, is changed and is not too small or too large. Therefore, from now on we evaluate and show the results by fixing the acceptable uncertainty at $20$ per cent.

\begin{figure}
    \centerline{
        \includegraphics[width=0.5\textwidth]{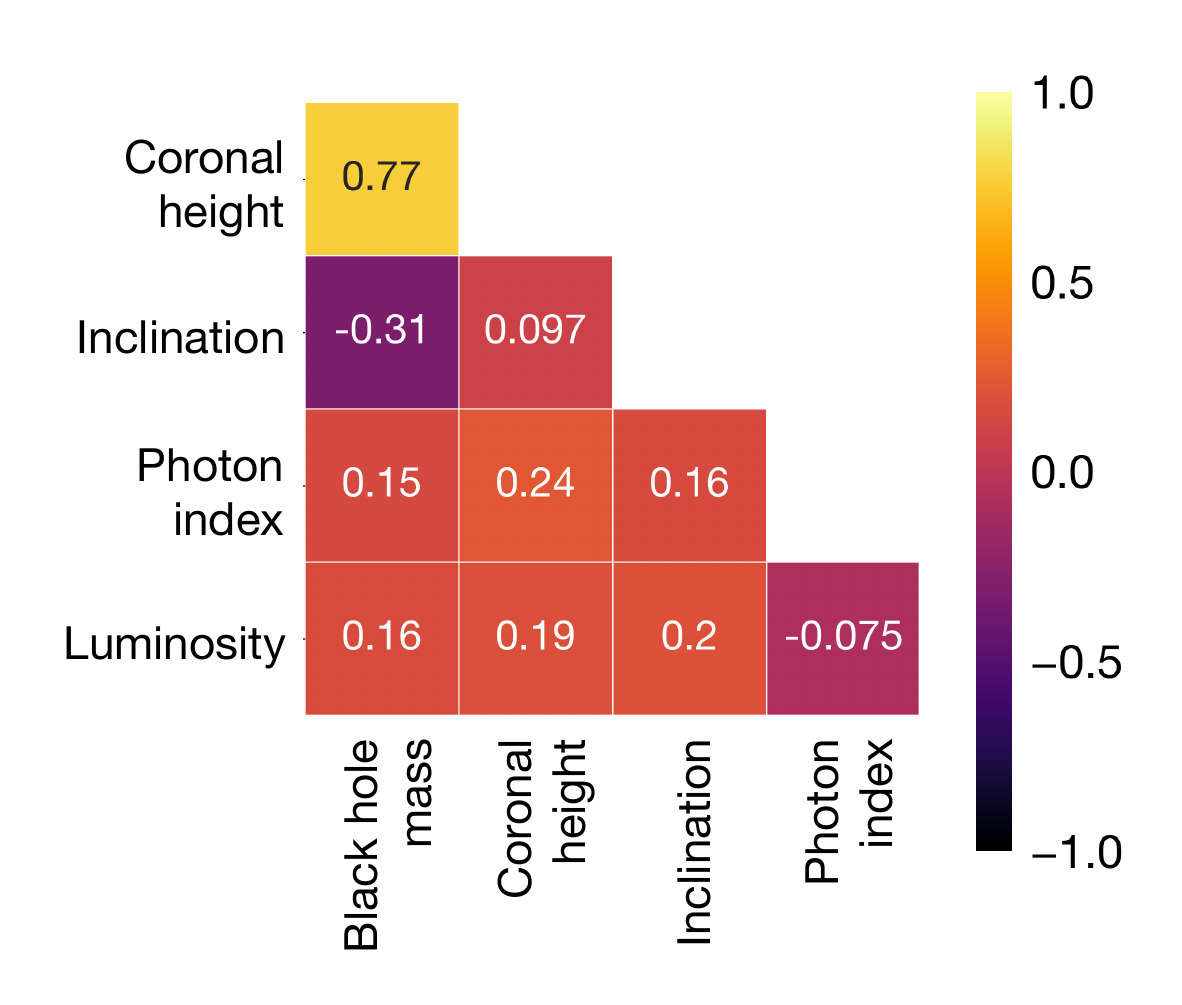}
        \put(-115,175){{\sc {\sc kynxilrev}}}
%\put(-131,134){$t_{\rm w}=200$~s}
    }
    % \vspace{-0.15cm}
    \centerline{
        \includegraphics[width=0.5\textwidth]{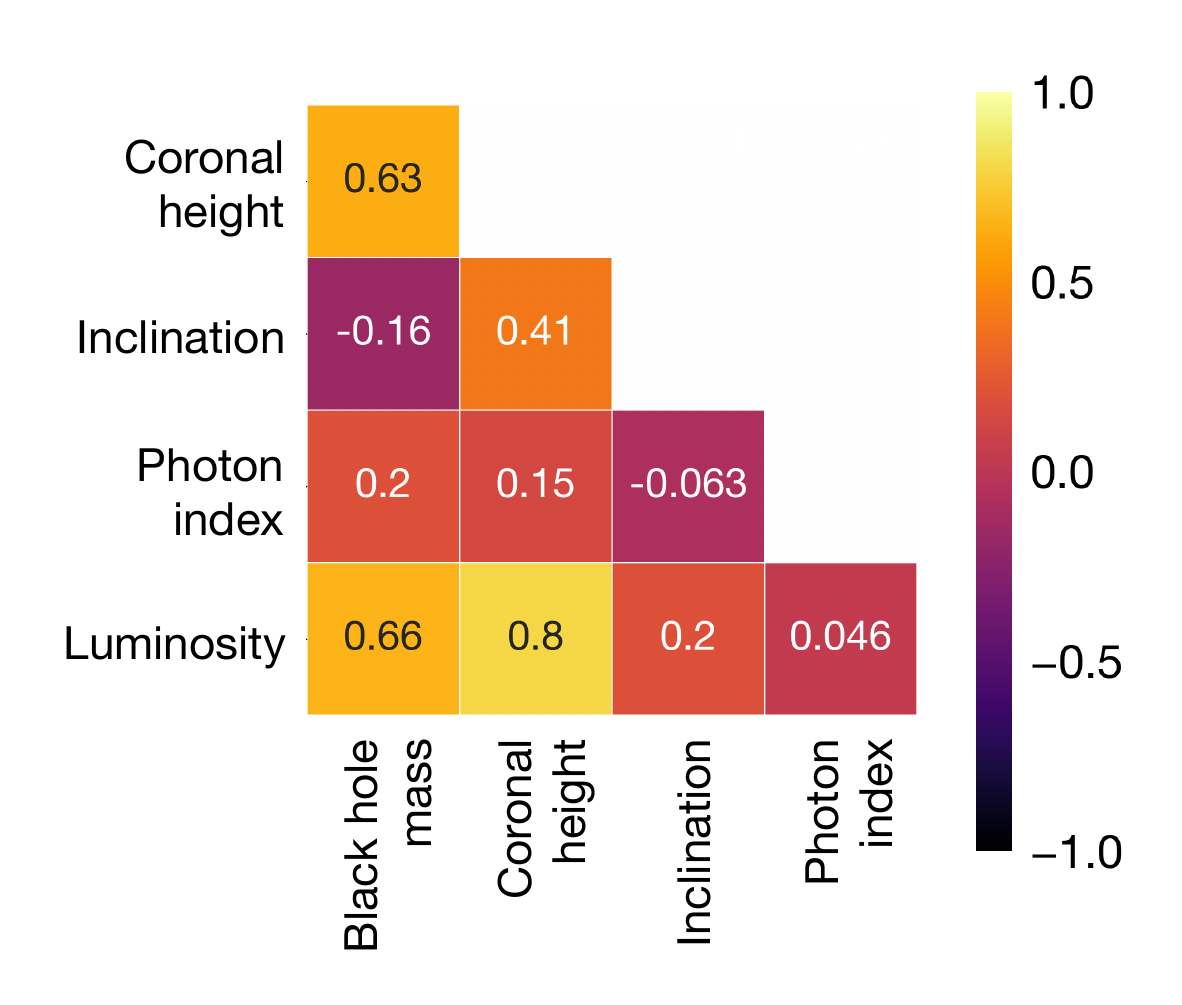}
        \put(-115,175){{\sc {\sc kynrefrev}}}
    }
    %\vspace{-0.2cm}
    \caption{Spearman's rank correlation coefficient ($r_{\rm s}$) between the obtained parameters constrained by the {\sc kynxilrev} (top panel) and {\sc kynrefrev} (bottom panel) models.}
    \label{fig:correlation}
\end{figure}

Overall correlations between each AGN parameter are presented in Fig.~\ref{fig:correlation}. Both {\sc kynxilrev} and {\sc kynrefrev} reveal a strong monotonic correlation between $h$ and ${M_{\rm BH}}$, with the Spearman's rank correlation coefficient $r_{\rm s} \sim 0.6$--0.8. The inclination is moderately correlated with the height ($r_{\rm s} = 0.41$) for the {\sc kynrefrev} model, while this correlation is not observed when using {\sc kynxilrev}. Notably, the luminosity also shows a high correlation with the coronal height only in the case of {\sc kynrefrev} model ($r_{\rm s} = 0.80$), while the {\sc kynxilrev} model does not reveal this correlation. An increasing trend of $L$ with $M_{\rm BH}$ is also seen only in {\sc kynrefrev} data.  

\begin{figure}
    \centerline{
        \includegraphics[width=0.35\textwidth]{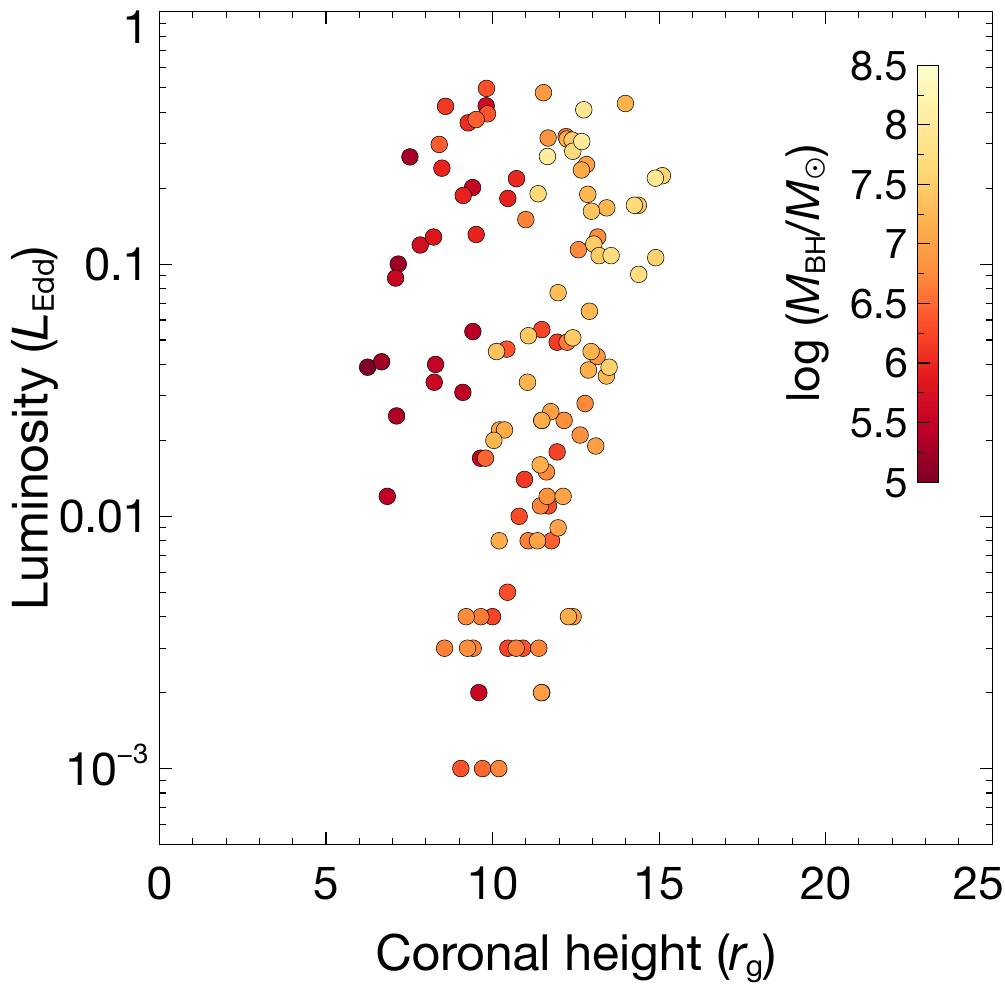}
        \put(-65,45){{\sc KYNXILREV}}
%\put(-131,134){$t_{\rm w}=200$~s}
    }
    % \vspace{-0.15cm}
    \centerline{
        \includegraphics[width=0.35\textwidth]{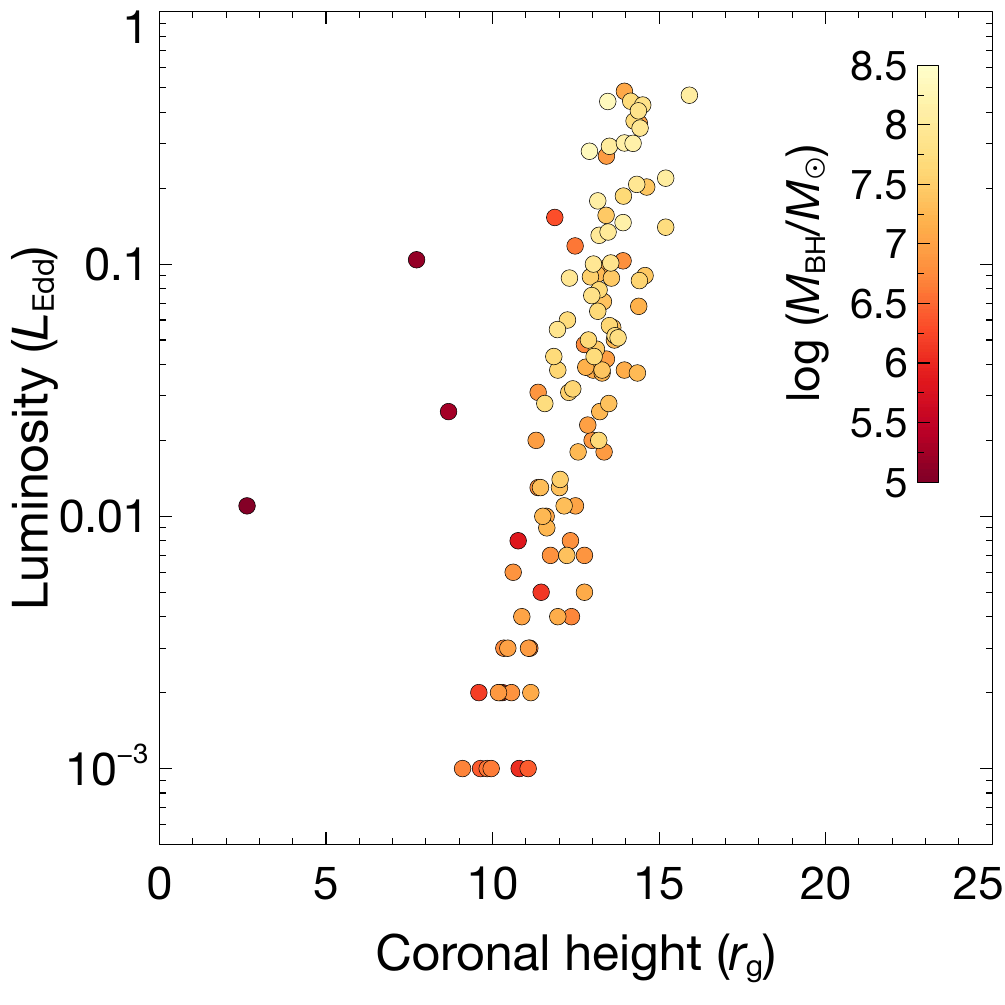}
        \put(-65,45){{\sc KYNREFREV}}
    }
    %\vspace{-0.2cm}
    \caption{Comparison of the $L$--$h$ relations derived by {\sc kynxilrev} (top panel) and {\sc kynrefrev} (bottom panel) models. Different colors indicate different $M_{\rm BH}$.}
    \label{fig:L-h}
\end{figure}

The distinct results of the $L$--$h$--$M_{\rm BH}$ relation obtained from both models are illustrated in Fig.~\ref{fig:L-h}. The mock data from {\sc kynxilrev} appear to be more scattered with the additional data of $M_{\rm BH} \lesssim 10^6 M_{\odot}$ compared to those from the {\sc kynrefrev}. These low mass data from the {\sc kynxilrev} model also correspond to relatively low source height and high luminosity, occupying the region in the parameter space where we find less solutions if using the {\sc kynrefrev} model. Comparing to Fig.~\ref{fig:height-mass}, these data are subjected to have high inclinations as well.

%\begin{figure}
% \includegraphics[width=\columnwidth]{Fig_01-3.png}
% \put(-70,343){$M_{\rm BH} = 10^6 M_{\odot}$}
% \put(-70,100){$M_{\rm BH} = 10^7 M_{\odot}$}
% \caption{Example of the lag-frequency spectra when varying the coronal height compared between {\sc kynxilrev} (dashed lines) and {\sc kynrefrev} (solid lines) and in two different black hole mass which are $M_{\rm BH} = 10^{6}$ and $10^{7} M_\odot$ (upper and lower panel). The other AGN parameters are set to be default that $a =$ 1, $i =$ 30, $\Gamma =$ 2 and $L =$ 0.001 $L_{\rm Edd}$.}
% \label{fig:lag-freq_spectrum}
%\end{figure}

\begin{figure}
    \centerline{
        \includegraphics[width=0.3\textwidth]{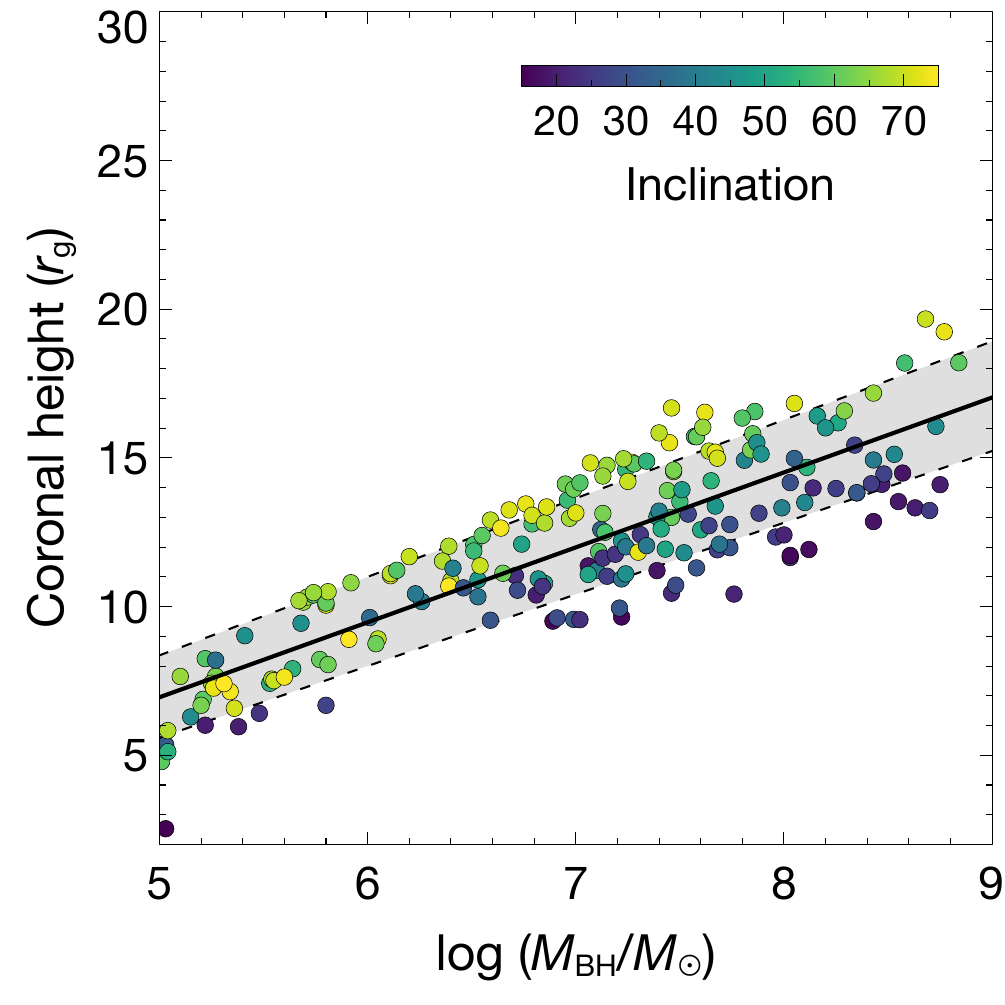}
        \put(-70,42){$\alpha = -5.65 \pm 0.81$}
        \put(-70,32){$\beta = 2.52 \pm 0.11$}
%\put(-131,134){$t_{\rm w}=200$~s}
    }
    % \vspace{-0.15cm}
    \centerline{
        \includegraphics[width=0.3\textwidth]{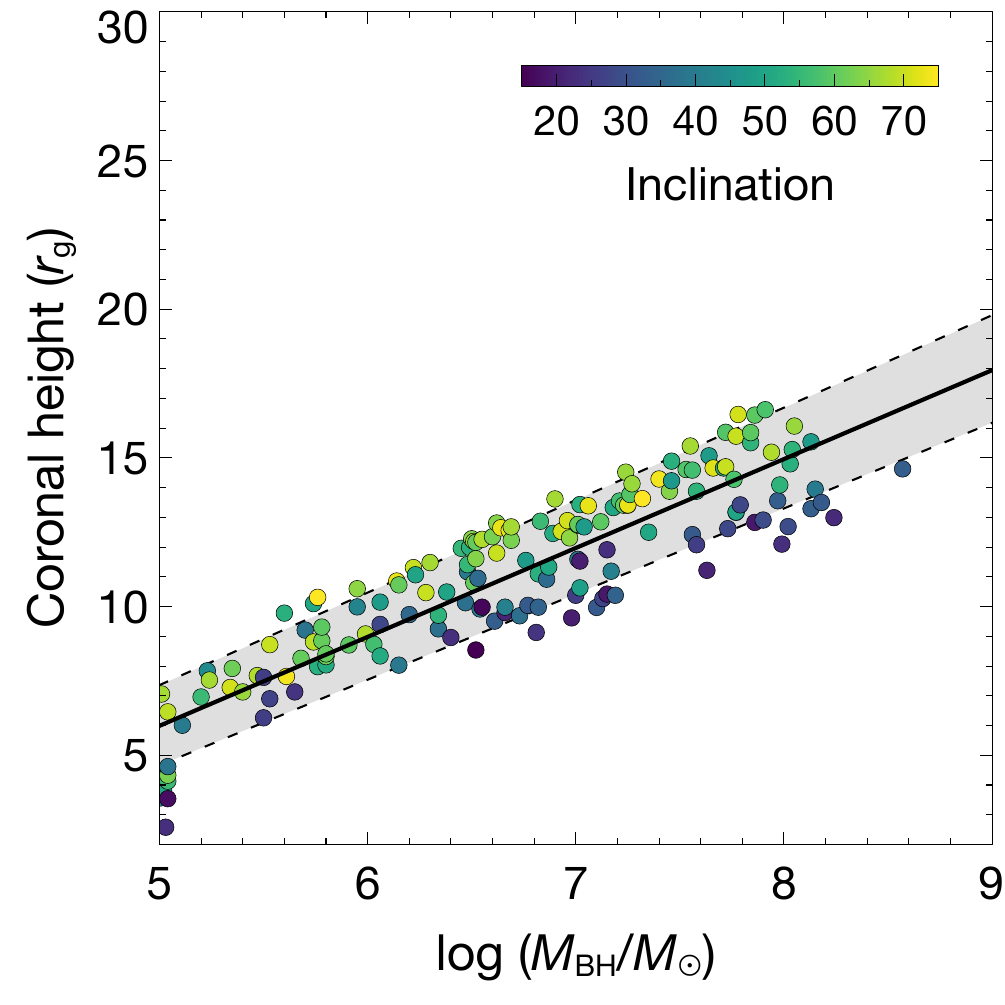}
        \put(-70,42){$\alpha = -8.96 \pm 0.78$}
        \put(-70,32){$\beta = 2.99 \pm 0.11$}
        %\put(-65,45){{\sc KYNREFREV}}
    }
     \centerline{
        \includegraphics[width=0.3\textwidth]{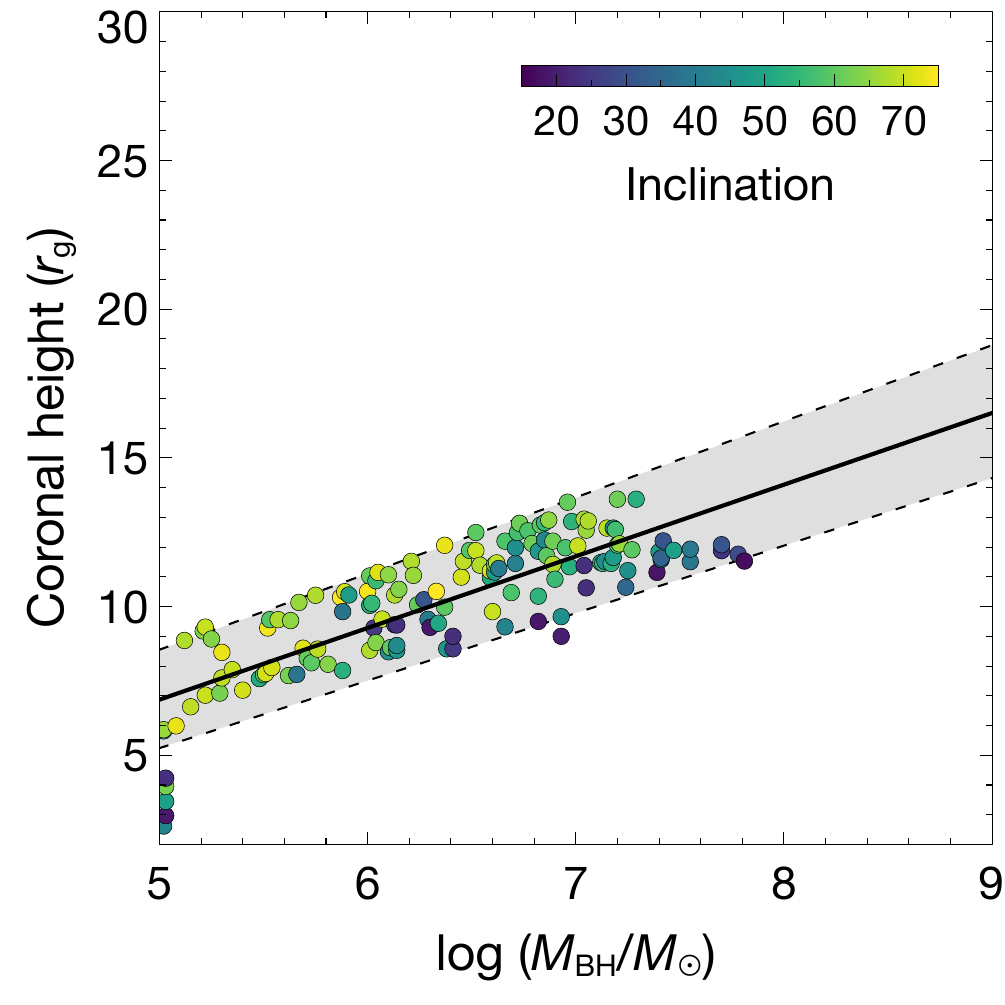}
        \put(-70,42){$\alpha = -5.18 \pm 0.93$}
        \put(-70,32){$\beta = 2.41 \pm 0.14$}
        %\put(-65,45){{\sc KYNREFREV}}
    }
     \centerline{
        \includegraphics[width=0.3\textwidth]{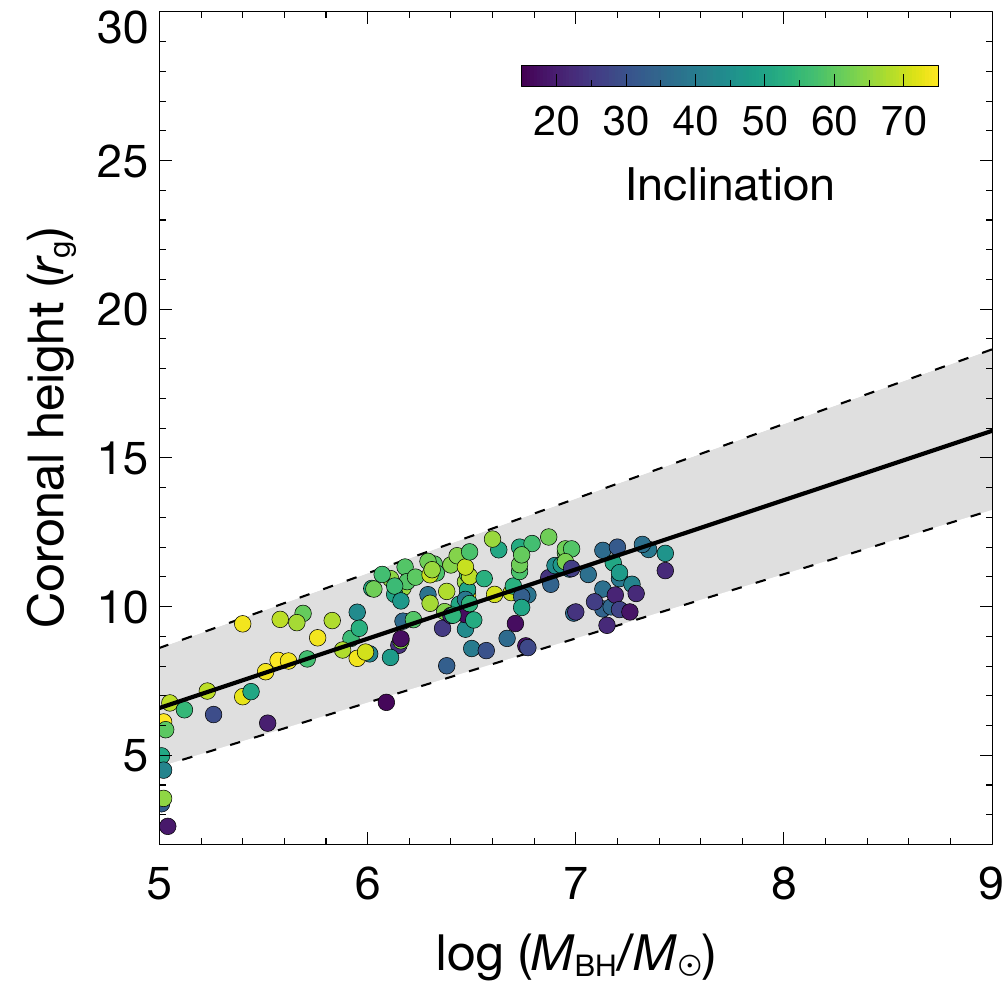}
        \put(-70,42){$\alpha = -5.06 \pm 1.12$}
        \put(-70,32){$\beta = 2.33 \pm 0.18$}
        %\put(-65,45){{\sc KYNREFREV}}
    }
    \vspace{-0.2cm}
    \caption{Obtained $h$--$M_{\rm BH}$ relations from the {\sc kynxilrev} model when the coefficient of the $\tau$--$M_{\rm BH}$ scaling relation is changed to be $-20$, $-10$, $+10$ and $+20$ per cent from the current trend, from top to bottom panels. The best-fit linear model is in the form of $h = \alpha + \beta \log (M_{\rm BH}/M_{\odot})$, yielding a $R^2$ of 0.71, 0.80, 0.66 and 0.58, for the fits from top to bottom panels, respectively.}
    \label{fig:model-eq}
\end{figure}

Furthermore, we investigate the effects when the slope of the observed lag-mass scaling relation (i.e. the coefficients on eq.~\ref{eq:lag-mass}) are deviated from the current trend. We adjust the slope of the $\tau$--$M_{\rm BH}$ profile to be $\pm 10$ and $\pm 20$ per cent from the recent one and generate new mock AGN samples that follow the new $\tau$--$M_{\rm BH}$ relation. The results are shown in Fig.~\ref{fig:model-eq}. It is clear that $h$ and $M_{\rm BH}$ is still correlated in all cases. A larger number of higher-mass AGN possessing larger coronal height is expected if the coefficient of the lag--mass relation decreases. In other words, more solutions with higher $h$ (up to $\sim 20~r_{\rm g}$) can be found only when the data lies in the trend that has smaller coefficients than the observed one. By assuming the relation in the form of $h = \alpha + \beta \log (M_{\rm BH}/M_{\odot})$, we see that $\alpha$ vary approximately between $-9$ to $-5$ while $\beta$ changes between 2.3 and 3. For a fixed mass, varying the slope of the lag-mass scaling law may represent the case when one looks at a particular AGN that exhibits different time lags as observed in different observations (see, e.g., \citealt{Alston2020} in the case of IRAS~13224--3809 where the lags and the inferred coronal heights are varied across different observations).

\section{Discussion and conclusion}

The X-ray reverberation lags in AGN were found to scale with the BH mass \citep{DeMarco2013, Kara2016,Hancock2022}. By investigating Fe-L lags, \cite{DeMarco2013} found that the lags roughly lie within the light crossing time for a distance of $6~r_{\rm g}$. \cite{Kara2016} analyzed the lags measured in the Fe-K band and found that the observed lags lie in the range of timescales corresponding to the light crossing distance of $\sim 1$--$9~r_{\rm{g}}$. These previous studies suggest that the distances associated with X-ray reverberation are small, consistent with the inner disc reflection framework. However, the observed lag amplitudes are usually smaller than the intrinsic value due to the dilution effects caused by the cross-contamination of primary and reflection flux in the energy bands of interest \citep[e.g.][]{Kara2013, Wilkins2013, Chainakun2015}. The true light travel distance then must be larger than those derived by converting the observed lags directly to the distance. Here, we follow the lag-mass scaling relation presented in \cite{DeMarco2013} and generate mock AGN samples for testing the consistency between the {\sc kynxilrev} and {\sc kynrefrev} in explaining these data. Both models suggest that the observed lags are mass-dependent and can be triggered by a compact lamp-post source lying within $h \lesssim 15~r_{\rm g}$.

The fact that the correlation coefficient between the lags and the mass is not intrinsically equal to 1 suggests that the lags must depend on other parameters as well. In addition to the black hole mass, it is likely that the observed lags also depend on the coronal height which is not the same in all AGN. The $\tau$--$h$ scaling relation is suggested either when using {\sc kynrefrev} or {\sc kynxilrev} model (Fig.~\ref{fig:lag-h_correlation} and eqs.~\ref{eq:lag-h-xil}--\ref{eq:lag-h-ref}). The slope of the $\tau$--$h$ fits from both models are comparable. Major differences between the {\sc kynxilrev} and {\sc kynrefrev} models are their underlying reflection spectra, {\sc xillver} and {\sc reflionx} respectively, where the reflection spectra employed by {\sc kynxilrev} were reported to exhibit a higher level of absorption particularly in the soft energy range of $\sim 0.3$--0.8~keV \citep{Garcia2013, Caballero2018}. Despite this, both models still consistently infer the monotonic correlation between the coronal height in the gravitational units and the central mass. In fact, the height and the mass are correlated such that they can also be used as another independent scaling law (Fig.~\ref{fig:height-mass} and eqs.~\ref{eq:h-mass-xil}--\ref{eq:h-mass-ref}).

The results suggest that, by analyzing the Fe-L lags alone, the black hole spin cannot be well constrained. There is no preferred value of the black hole spin since the derived solutions from both models can match the mass-scaling law whether the black hole is non-, moderately-, or maximally-spinning. This is probably because the lag-frequency spectrum has a subtle change with the spin \citep{Cackett2014}. Even if this is the case, the {\sc kynxilrev} model seems to provide more solution towards lower $h$ and lower $M_{\rm BH}$ especially for a low-spin AGN. 

Since the {\sc kynxilrev} model has more dilution, one would expect that this model requires longer intrinsic lags and higher coronal heights to explain the observed data. However, {\sc kynxilrev} allows smaller values of height compared to {\sc kynrefrev}. This is because, for a fixed coronal height, the reverberation lag from {\sc kynxilrev} is smaller than that from {\sc kynrefrev}, but their phase wrapping occurs at the same frequency (Fig.~\ref{fig:lag-freq_spectrum}). When the {\sc kynxilrev} model requires a higher source height to explain the lags, the phase-wrapping also shifts to lower frequencies than that from the {\sc kynrefrev} model, thus it no longer matches the mass scaling relation.

\cite{Alston2020} simultaneously fit the lag-frequency spectra in 16 {\it XMM-Newton} observations of IRAS~13224--3809 using the {\sc kynrefrev} model. While the variation of coronal height in IRAS~13224–3809 was found to be $h \sim 6$--$20~r_{\rm g}$, the majority of the obtained heights were found to be $h \lesssim 13~r_{\rm g}$. Note that the $h$--$M_{\rm BH}$ relation obtained when the slope of the lag-mass scaling law is varied can show how much the coronal height can change for a fixed mass. Given the IRAS~13224–3809 mass of $\log(M_{\rm BH}/M_{\odot}) \sim 6.38$ \citep{Alston2020}, their upper limit of the height is roughly consistent with our possible heights at that particular mass. \cite{Caballero2020} investigated the combined spectral-timing data of IRAS~13224–3809 and found that, for the maximally spinning case, the source height could vary between $\sim 3$--$10~r_{\rm g}$. Recently, \cite{Mankatwit2023} utilizes a random forest regressor machine-learning model to investigate the X-ray reverberation lags due to the lamp-post corona in IRAS~13224–3809 and 1H~0707–495 using the response functions from the {\sc kynxilrev} model. Their constrained heights were also varied between $5$--$18~r_{\rm g}$. Following the current lag-mass scaling relation, it is rare that the AGN would occupy the lamp-post corona at a distance larger than $\sim 15~r_{\rm g}$. Otherwise, the coefficients of the mass scaling law needs to deviate from the current trend by about $20$~per cent (Fig.~\ref{fig:model-eq}).

% ///////-- Okay --////////
\cite{DeMarco2013} also investigated whether the lag timescales show some dependence on the source luminosity. They found a less significant correlation between the Fe-L lags and either the 2--10 keV or bolometric luminosity. Here, the significantly strong or even moderate correlations between the lags and the $L$ cannot be observed. In fact, our results suggest that we should not expect to see the correlations between the lags and either $\Gamma$ or $i$ too (Fig.~\ref{fig:lag_corelation}). This means that $L$, $\Gamma$ and $i$ are not the key parameters that could place constraints on the lags and the disc-corona geometry in AGN. By using the Fe-K lags instead, \cite{Kara2016} found that the coronal height is correlated with the Eddington ratio, with the Spearman’s rank correlation coefficient $r_{\rm s} = 0.61$. Here, we found a strong correlation between the source height and the source luminosity ($r_{\rm s}= 0.80$), but only when using the {\sc kynrefrev} model (Fig. \ref{fig:L-h}). This leads us to question if there really is a true correlation between the source geometry and luminosity, or if it is dependent on the choice of the reflection and reverberation models.

The extended study from \cite{Mallick2021} included more samples of low-mass AGNs ($M_{\rm BH} < 3 \times 10^6 \:M_{\odot}$) to the mass-scaling investigation. The equation of the lag-mass scaling is suggested to be comparable to those of \cite{DeMarco2013}, confirming that the soft lag amplitude in higher-mass AGN can be scaled with the mass to infer the lags in lower-mass AGN. Furthermore, they found that the corona extends at an average height of $\sim 10~r_{\rm g}$ on the symmetry axis. Our results suggest that the coronal height at $\sim 10~r_{\rm g}$ can potentially be found from low-mass to high-mass end of AGN (Fig.~\ref{fig:height-mass}) either using {\sc kynxilrev} or {\sc kynrefrev}. The deviation of the lag-mass equations comparing between \cite{DeMarco2013} and \cite{Mallick2021} should be within the $\pm 20$ per cent of the lag variations allowed in this work (see Fig.~\ref{fig:model-eq}), so our results should still be representative either for the samples of low- or high-mass AGN in previous literature.

Our obtained $h$--$M_{\rm BH}$ relation suggests that the corona in lower-mass AGN tends to be more compact than the corona in higher-mass AGN. Although timing analysis under the lamp-post assumption successfully determines the coronal geometry \citep{Caballero2018}, it may prefer higher source heights compared to the method using the time-averaged spectra \citep{Alston2020, Chainakun2022a, Jiang2022}. The real corona should extend and different parts may vary differently. The extended corona was suggested in many studies to have an ability to influence the time-lag spectra that can explain the data \citep{Wilkins2016, Chainakun2017, Hancock2023, Lucchini2023}. In this case, the same corona geometry can produce different reverberation lag amplitudes by varying its properties such as the optical depth \citep{Chainakun2019} and the propagating fluctuations inside the corona \citep{Wilkins2016}, or by assuming different disc geometry \citep{Taylor2018, Kawamura2023}. Adjusting these are not possible for the current {\sc kynxilrev} and {\sc kynrefrev} models.

Futhermore, \cite{Wilkins2020} developed a modified {\sc xillver} model referred to as {\sc xillverRR} to study the returning radiation that can cause extra reverberation delays due to secondary or even higher order reflections. The results show that, for the black hole with the spin $a = 0.998$ and the corona at $h = 5 \;r_{\rm g}$, almost 40 per cent of the photons will experience the returning process. The returning fraction will decrease as the X-ray source height increases, with the crucial point at $h=10~r_{\rm g}$ where the photons are more likely to travel directly to the observer. Both {\sc kynxilrev} and {\sc kynrefrev} do not take into account the extra time delays due to the returning radiation. This can lead to an uncertainty in the measurement of the coronal height especially at $h \lesssim 5~r_{\rm g}$ where the returning radiation should play an important role. As for that, the obtained source height at $h \lesssim 5~r_{\rm g}$ could, in fact, be smaller if we consider the effect of the returning radiation. However, this effect will similarly apply to both models, by mainly adding extra light-travel time so it should not change the comparative results between both models. 

Furthermore, the effects of non-uniform orbiting clouds that introduce the covering fraction \citep{Hancock2022} and extra dilutions that may occur due to environmental absorption \citep{Parker2021} are not taken into account in calculating the lags. A recent study by \cite{Jaiswal2023} suggested that the effect of the scattering of the disc emission by the broad line region on the time delay is quite similar to the effect of increasing the X-ray source height. Again, the non-quantitative comparison of {\sc kynxilrev} and {\sc kynrefrev} should still be valid since all these effects will similarly apply over both models.

In conclusion, the maximum coronal height inferred by {\sc kynrefrev} and {\sc kynxilrev} models that assumes a lamp-post model for a distant observer is likely limited at $h \lesssim 15$--20$~r_{\rm g}$. This should be true for any newly-discovered AGN if their lags and mass follows the current scaling-relation trend. The average source height at $h \sim 10~r_{\rm g}$ can be found either in low or high mass AGN. Our results reveal that there is a high chance that the lower-mass AGN has a more compact corona located at a lower height on the symmetry axis. As for that, the $h$--$M_{\rm BH}$ scaling relation may be valid whether the coronal height is implied using {\sc kynrefrev} or {\sc kynxilrev} model. However, the differences of both models in explaining the X-ray timing data are clearly seen. There is an inconsistency that the {\sc kynxilrev} suggests a greater number of possible solutions with low $M_{\rm BH}$ and low $h$ than the {\sc kynrefrev}, especially for the low spinning black hole. Last but not least, the {\sc kynrefrev} models suggest a moderate monotonic correlation between the luminosity and the coronal height while the {\sc kynxilrev} does not. Therefore, when analysing a large number of data or observations, the derived parameter correlations may be model dependent.

\section*{Acknowledgements}

This work was supported by the NSRF via the Program Management Unit for Human Resources \& Institutional Development, Research and Innovation (grant number B16F640076). KK acknowledges the support from the Institute for the Promotion of Teaching Science and Technology (IPST) of Thailand. PC thanks for the financial support from Suranaree University of Technology (grant number 179349). We thank the anonymous referee for their comments which improved the clarity of the paper.

\section*{Data availability}
The reverberation models including the {\sc kynxilrev} and {\sc kynrefrev} used in the simulation process are under the {\sc kynreverb} package, available in \url{https://projects.asu.cas.cz/stronggravity/kynreverb}. The simulated and analysed data underlying this article will be shared upon reasonable request to the corresponding author.


\bibliographystyle{mnras}

\appendix
\section{Variable inner-disc radius}
\label{sec:a1}

\begin{figure}
 \includegraphics[width=\columnwidth]{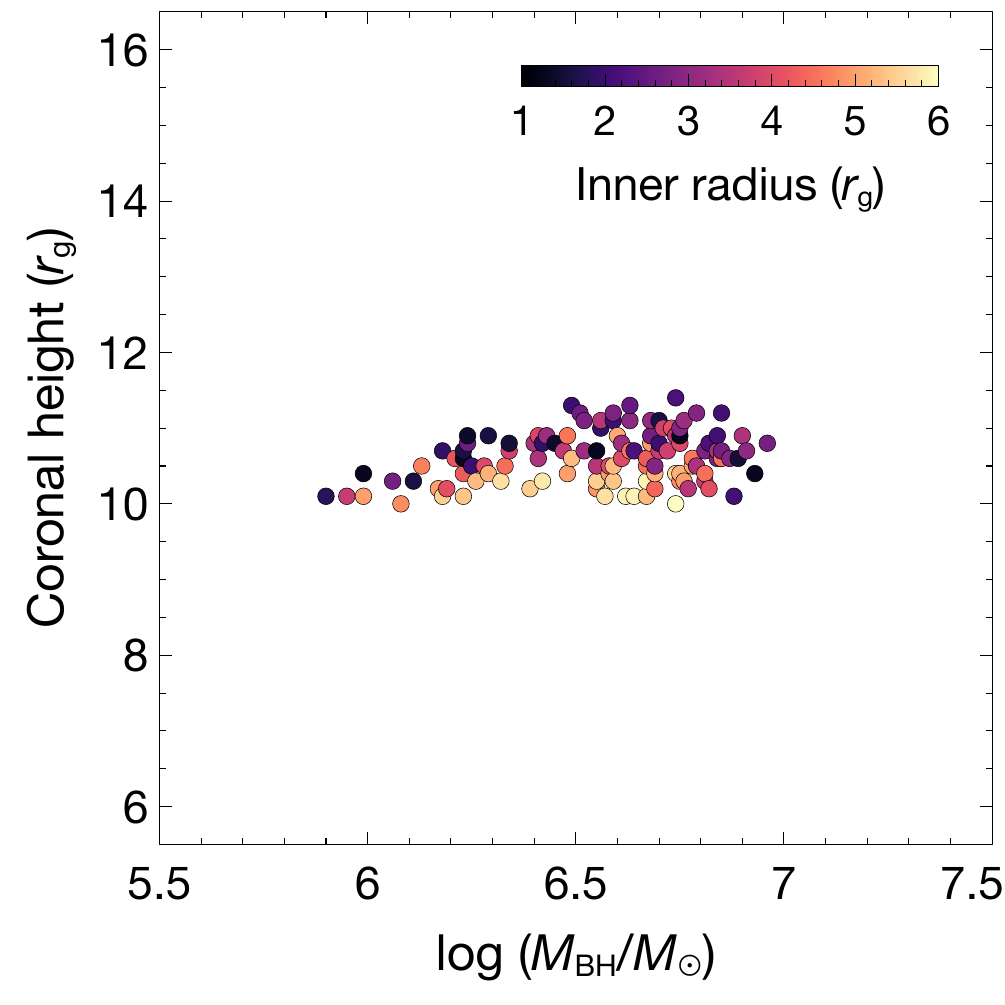}
 \vspace{0 cm}
 \caption{The $h$--$M_{\rm BH}$ relation from {\sc kynxilrev} when $a = 1$ and the inner-disc radius is allowed to vary randomly between ISCO and $6~r_{\rm g}$. We fix $i = 45^{\circ}$, $\Gamma = 2.0$ and $L/L_{\rm Edd} = 0.1$ in this illustration. For a variable inner radius, the model provides more possible solutions with lower source heights for a given mass.} \label{fig:r_in-test}
\end{figure}

We further investigate the case of a variable inner-disc radius. We fix $a = 1$, $i = 45^{\circ}$, $\Gamma = 2.0$ and $L/L_{\rm Edd} = 0.1$ while randomly vary the truncation radius of the disc. Other parameters including $M_{\rm BH}$ and $h$ are allowed to vary in the range given in Table~\ref{tab:parameter}. The result is shown in Fig.~\ref{fig:r_in-test}. While the lag is driven mostly by the distance between the corona and the disc, a larger truncation radius results in a larger distance for the coronal photons to travel to the inner edge of the disc, hence producing a longer lag. Therefore, if the disc is allowed to be truncated, there is a chance to get lower values of the height for a given mass, as can be seen in a clear separation of color in the $h$--$M_{\rm BH}$ relation. There is only a minority of the data that show higher coronal heights for a truncated disc compared to when the disc is fixed to ISCO. This inconsistency should arise due to the fact that we allow some threshold levels of the lag amplitude. However, the accreting supermassive black holes in AGN are mostly rapidly spinning and we do not expect the AGN disc to be significantly truncated since the broad emission lines are often observed \cite{Reynolds2021}.

%It can be seen that the inner radius can play a role in changing the reverberation lag that leads to lower values of the height for a given mass. This is because the lag is driven mostly by the distance between the corona and the disc, and a larger truncation radius results in a larger distance for the coronal photons to travel to the inner edge of the disc, hence producing a longer lag. It then requires a lower source height compared to the case when the inner radius is fixed at ISCO, leading to a shallower slope for the $h$--$M_{\rm BH}$ relation. However, the accreting supermassive black holes at the centre of AGN are mostly rapidly spinning and we do not expect the AGN disc to be significantly truncated since the broad emission lines are often observed \cite{Reynolds2021}.

\section{Sources of Uncertainty}
\label{sec:a2}

\begin{figure}
 \includegraphics[width=\columnwidth]{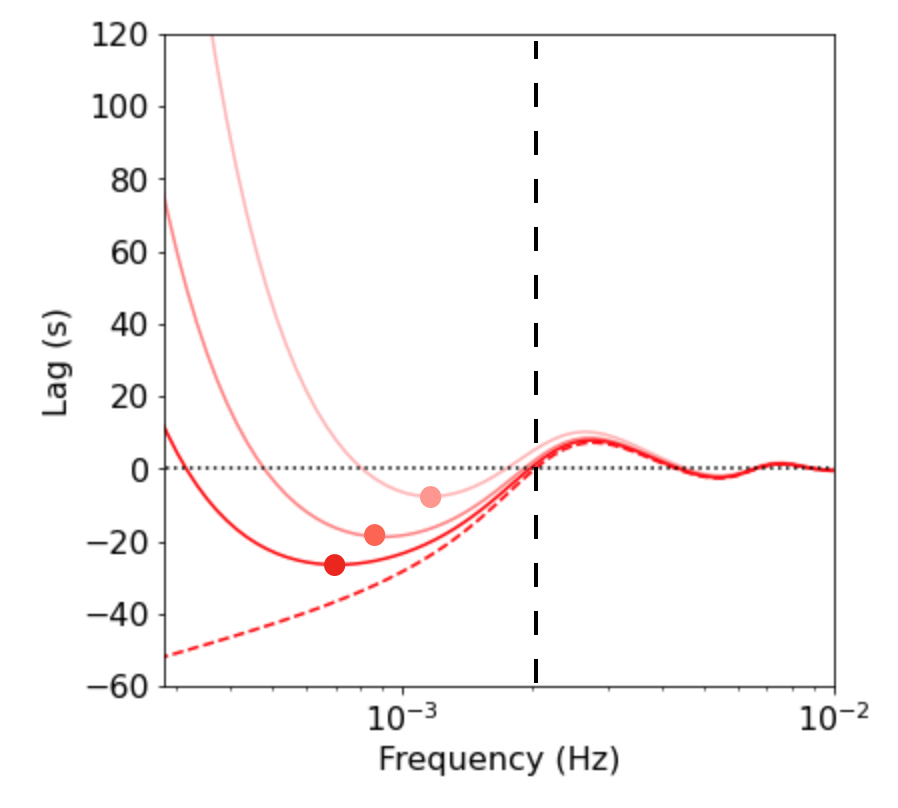}
 \put(-78,200){{$h = 20~r_{\rm g}$}}
 \put(-78,188){{$i = 30^{\circ}$}}
 \put(-78,176){{$\Gamma = 2$}}
 \put(-78,164){{$L = 0.001L_{\rm Edd}$}}
 \put(-78,152){{$M_{\rm BH} = 10^{6}M_{\odot}$}}
 \vspace{0 cm}
 \caption{Simulated lag-frequency spectra from {\sc kynrefrev} (red dahsed line) where the vertical dashed line identifies the phase-wrapping frequency. Red solid lines represent the lag-frequency spectra that also include positive propagating-fluctuation lags modelled as a power-law with different normalizations. Each circle locates the minimum point of each profile, showing that amplitude of the reverberation lag and the associating frequency at the minimum point can be affected by the propagating-fluctuation process leading to an uncertainty in determining their true values.} \label{fig:uncertainty}
\end{figure}

We note an effect of the positive hard lags due to the inward propagation of fluctuations in the disc \citep{Arevalo2006} that may contribute to the uncertainty in our results. The observed reverberation lags further away from the phase wrapping towards lower frequencies will be likely more affected by the propagating-fluctuation lags that operate on relatively long timescales. In Fig.~\ref{fig:uncertainty} we show how this competing process can affect the reverberation lags at frequencies where we expect to see the reverberation lags. A full modelling of the disc-propagating fluctuations could be an option but it may lead to a significantly larger set of parameters. In Fig.~\ref{fig:uncertainty}, the amplitude and frequency of the reverberation lag at the minimum point move closer to the phase wrapping frequency if the disc propagating-fluctuation signals become stronger. Also, if the coronal height increases, the phase wrapping point will move towards lower frequencies. The reason why large distances (e.g. $h \gtrsim 20~r_{\rm g}$) between the lamp-post source and the disc do not explain the observations is probably because their phase wrapping frequencies are too low. Unless the true positive lags are known, there is always uncertainty in determining the exact frequency and amplitude of the minimum point in the negative-lag profiles. Nevertheless, these uncertainties may be compensated more or less by 1) allowing some threshold levels of the lag amplitude and 2) investigating the cases when the mass scaling relation deviates from the current observed trend (Fig.~\ref{fig:model-eq}). Therefore, the results here should still be reliable and representative of the overall parameter relations.

\bsp	% typesetting comment
\label{lastpage}
\end{document}